\documentclass{article}
\normalsize
\usepackage{multicol}

\title{Example of an explicit function for confining classical Yang-Mills
fields with quantum fluctuations in the path integral scheme}
\author{Kimichika~Fukushima$^{1}$ and Hikaru~Sato$^{2}$\\
\\
$^{1}$ Advanced Reactor Engineering Department,\\
Toshiba Corporation,\\
8, Shinsugita-cho, Isogo-ku, Yokohama 235-8523, Japan\\
$^{2}$ Emeritus, Department of Physics, Hyogo University of Education,\\
Yashiro-cho, Kato-shi, Hyogo 673-1494, Japan}

\date{             }

\begin{document}

\maketitle

This article reports an explicit function form for confining classical Yang-Mills vector potentials and quantum fluctuations around the classical field. The classical vector potential, which is composed of a confining localized function and an unlocalized function, satisfies the classical Yang-Mills equation. The confining localized function contributes to the Wilson loop, while the unlocalized function makes no contribution to this loop. The confining linear potential between a heavy fermion and antifermion is due to (1) the Lie algebra and (2) the form of the confining localized function which has opposite signs at the positions of the particle and antiparticle along the Wilson loop in the time direction. Some classical confining parts of vector potentials also change sign on inversion of the coordinates of the axis perpendicular to the axis joining the two particles. The localized parts of the vector potentials are squeezed around the axis connecting the two particles, and the string tension of the confining linear potential is derived. Quantum fluctuations are formulated using a field expression in terms of
local
basis functions
in real spacetime.
The quantum path integral gives the Coulomb potential between the two particles in addition to the linear potential due to the classical fields.
\par
\verb+ +
\par

\begin{multicols}{2}

\section{Introduction}

Quantum field theory \cite{PRam,Aber} has been successfully applied to various perturbative and non-perturbative phenomena [3-12]. Some examples of non-perturbative phenomena are fermion confinement and superconductivity. The present authors have studied some of these problems in quasi-low-dimensional cases [13-15]. Non-perturbative methods are required when the expansion series with respect to a coupling constant converges slowly (the non-Abelian gauge field case is often involved). In our earlier papers, we have presented a formulation to expand fields in terms of
local
basis functions
in real spacetime
\cite{Fuku84}. Another method, which is rather different from our method, was also reported \cite{BenMS}. Note, if the zeroth-order approximation of
a classical field
(as a possible vacuum)
provides a good description of the phenomenon in question, it is reasonable to expect that the inclusion of quantum fluctuations beyond the zeroth-order approximation will also provide a good description of the phenomenon \cite{PRam}. This article presents an approach to fermion confinement in Yang-Mills fields, along this line of thought.
\par
Explicit forms for confining vector potentials of Yang-Mills fields have, so far, not been fully understood. In this paper, we present an example of an analytic classical vector potential composed of a localized function and an unlocalized function in the center-of-mass frame of a heavy fermion-antifermion pair. The total field, shifted from the zero field, satisfies the classical equations of motion. The localized and unlocalized functions have different behaviors. The localized function rapidly decays in the region away from the axis connecting the two particles, whereas the unlocalized function decreases slowly. Additionally, the localized function has a finite value at a specific point, whereas the unlocalized function vanishes because of cancellation between terms with opposite signs. The localized function of the classical field configuration results in the Wilson loop with a confining potential, whereas the unlocalized function does not contribute to the Wilson loop because of the cancellation between terms. The confinement is due to a property of the trace of matrix polynomials in Lie algebra, which are quite different from the Abelian case. The existence and stability of this classical configuration is determined from the action, which leads to the classical equations of motion, and the Wilson loop, which describes energy lowering in the confining phase as compared to the Coulomb phase. The classical vector potentials, we have investigated, have not been mentioned in other literature \cite{Act79}.
\par
The confining parts of the classical vector potentials are squeezed around the axis between a fermion and antifermion pair, and some change sign on inversion of coordinates of the axis normal to the axis joining the two particles.
A
parameter
in the classical vector potential
is determined using the
(renormalization) scale-invariant energy relation
in the pair creation process. The classical configuration treated here is
an ${\rm SU}(2)$ gauge field
embedded in ${\rm SU}(N)$,
but the quantum fluctuations are described by an ${\rm SU}(N)$ gauge field (such as ${\rm SU}(3)$), so that the derived results are for a gauge field like ${\rm SU}(N)$.
\par
In order to calculate quantum fluctuations around the classical field, we present a formulation for expressing fields in terms of
local
basis functions
in real spacetime
in the path integral scheme. We also show analytically that diagonalized quadratic terms have positive eigenvalues under a non-periodic boundary condition (zero eigenvalues result from a periodic boundary condition). We consider that real spacetime in four dimensions is open, and has a non-periodic boundary condition. The total potential is the sum of the linear potential due to the classical field and the Coulomb potential due to the quantum fluctuations around it. The classical field
gives
rise to a local mass (energy gap) term for quantum fluctuations of the Yang-Mills field in the action.
\par
A comparison can be made between the new classical solution presented in this paper and other approaches. Classical solutions were reviewed in literature \cite{Act79}, and subsequent some classical solutions \cite{VDzh} may exhibit confining linear-like (non-exact linear) potentials between the fermion particle and antiparticle. Quantum evaluations may be anticipated for previous classical solutions. Concerning a dual superconductor model, matrices with Abelian properties decomposed using an Abelian projection from the non-Abelian matrix resemble the original non-Abelian matrix. This model may reveal the condensation of magnetic monopoles into a superconducting state, causing a linear potential due to the squeezing of the electric field. The quantum confinement mechanism is currently a black box, meaning that its exact mechanism is unknown \cite{MShi}. A review on the dual superconductor model explained that the model would not confine the gauge field \cite{GRip}. It is also known that the gauge field wave, whose energy is less than the superconducting energy gap (transition temperature), propagates through real solid state (Abelian) superconductors \cite{CKit,MTin}. The lattice gauge theory has many advantages, and the confinement mechanism, which is considered to be black box, is expected to be understood. The string tension, computed using the lattice gauge theory, decreased from the original computed value \cite{HRot}. In contrast, our new solution has a classical explicit form of a function revealing a linear confinement potential, and the Coulomb potential is derived from quantum fluctuations around the classical field. Our solution has not so far been reported in literature and is quite different from other approaches. It appears that the quantum field has a local mass. Our confinement mechanism originates from the trace of the polynomials of the non-Abelian Lie matrix combined with a
solitonlike
localized function. As will be shown in detail in Subsection 2.3, the string tension derived from our solution is to be compared with the lattice gauge theory results, indicating that our result is compatible with the expected value dictated by the quantum chromodynamics (QCD).
\par
This article is organized as follows: Section 2 presents a set of classical vector potentials with confining properties. We also provide an expression for the string tension and the physical interpretation of classical vector potentials. Section 3 describes a formulation of fields expanded in terms of real spacetime basis functions for the evaluation of quantum fluctuations. Subsequently, Section 4 presents an analysis of quantum fluctuations around the classical configuration using this real spacetime basis set. Section 5 lists our conclusions.
\par

\section{Classical vector potentials of Yang-Mills fields}

\subsection{Classical field configuration for confinement}

A non-Abelian gauge field is represented by
\begin{eqnarray}
A_{\mu}(x)=A_{\mu}(t,x,y,z)
=A_{\mu}(t,{\bf x})=
\sum_{a}A_{\mu}^{a}(x)T^{a},
\end{eqnarray}
where $(t,{\bf x})=(t,x,y,z)$ and matrices $T^{a}$
generate
a Lie algebra
\begin{eqnarray}
[T^{a},T^{b}]=\sum_{c}if^{abc}T^{c}.
\end{eqnarray}
The indices $a$, $b$ and $c$ above run from 1 to $k_{\rm D}$, where $k_{\rm D}$ is the dimension of the Lie algebra, and $f^{abc}$ is the structure constant. In this paper, we work with ${\rm SU}(N)$ gauge fields (such as ${\rm SU}(3)$) as an example. The Lagrangian density for non-Abelian gauge fields in the case of the Feynman gauge in Euclidean spacetime is
\begin{eqnarray}
{\cal L}=\frac{1}{4} \sum_{a}F^{a}_{\mu\nu}F^{a}_{\mu\nu}
+\frac{1}{2}\sum_{a}(\partial_{\mu} A^{a}_{\mu})^{2},
\end{eqnarray}
where, setting $x_{0}=t$ we have the Euclidean spacetime metric 
$x_{\mu}x_{\mu}=x_{0}x_{0}+x_{1}x_{1}+x_{2}x_{2}+x_{3}x_{3}$,
rather than
$x_{\mu}x_{\mu}=-x_{0}x_{0}-x_{1}x_{1}-x_{2}x_{2}-x_{3}x_{3}$.
In the following, $t$, $x$, $y$ and $z$ are the dimensionless Euclidean time-space coordinates. When higher-order terms are considered, and if necessary, we add the following Feynman-Faddeev-Popov ghost terms, described with Grassmann fields $\omega^{a}$,
\begin{eqnarray}
{\cal L_{\rm FFP}}=
\sum_{a} \partial_{\mu}\omega^{*a}\partial_{\mu}\omega^{a}
-g\sum_{a,b,c}f^{abc}\partial_{\mu}\omega^{*a}A^{c}_{\mu}\omega^{b},
\end{eqnarray}
where the coupling constant is denoted as $g$. The action without the ghosts (in the classical case, it may be said that the action is defined around the ghost with a value of zero) is
\begin{eqnarray}
\nonumber
S=\int d^4x {\cal L}=\frac{1}{2}\sum_{a} \int d^4x 
(\partial_{\nu} A^{a}_{\mu}\partial_{\nu}A^{a}_{\mu})
\end{eqnarray}
\begin{eqnarray}
\nonumber
-\frac{1}{2} \sum_{a} \int d^4x 
(\partial_{\nu}(A^{a}_{\mu}\partial_{\mu}A^{a}_{\nu}
- A_{\nu}^{a}\partial_{\mu}A^{a}_{\mu} ))
\end{eqnarray}
\begin{eqnarray}
\nonumber
-g \sum_{a,b,c} f^{abc} \int d^4x
(A^{b}_{\mu}A^{c}_{\nu}\partial_{\mu}A^{a}_{\nu})
\end{eqnarray}
\begin{eqnarray}
+\frac{g^{2}}{4} \sum_{a,b,c,d,e} f^{abc}f^{ade} \int d^4x 
(A^{b}_{\mu}A^{c}_{\nu}A^{d}_{\mu}A^{e}_{\nu}),
\end{eqnarray}
where the four-dimensional volume element is given by
$d^{4}x=dtdxdydz$.
By rewriting the term above as
\begin{eqnarray}
S_{(2)}=-\frac{1}{2} \sum_{a} \int d^4x 
(\partial_{\nu}(A^{a}_{\mu}\partial_{\mu}A^{a}_{\nu}
- A_{\nu}^{a}\partial_{\mu}A^{a}_{\mu} )),
\end{eqnarray}
in terms of the surface integration, it is apparent this surface term can be neglected.
\par
We consider the case where a heavy fermion-antifermion is created at the origin of spacetime coordinates. Quantum fluctuations are defined, using the path integral around a classical configuration, which describes the zero field and satisfies the classical equations of motion. In contrast, classical Yang-Mills vector potentials $A^{a}_{\rm (C)\mu}$ with the index ${\rm (C)}$, which are treated by the present scheme, are shifted from $A^{a}_{\rm (C)\mu}=0$ to a non-zero field. This classical configuration
belongs
to a subgroup ${\rm SU}(2)$ of ${\rm SU}(N)$ (such as ${\rm SU}(3)$), but quantum fluctuations around this classical configuration are described by the ${\rm SU}(N)$ gauge field. Then, the total field configuration is given by the ${\rm SU}(N)$ gauge field and satisfies the classical equations of motion. The classical field configuration we have treated consists of two different functions
\begin{eqnarray}
A^{a}_{\rm (C)\mu}(x)=A^{a}_{{\rm (CL)}\mu}(x)+A^{a}_{{\rm (CU)}\mu}(x),
\end{eqnarray}
where $A^{a}_{{\rm (CL)}\mu}$ with the index ${\rm (CL)}$ is a localized field, which results in confinement and represents a specific solitonlike topological object. The other function $A^{a}_{{\rm (CU)}\mu}$ with the index ${\rm (CU)}$ is an unlocalized field, which does not contribute to the Wilson loop. The total classical field above is an embedded field, and we note that the confining property is caused by the characteristics of the Lie algebra in the action. This confining property is not seen in Abelian gauge fields without the Lie algebra in the action. The possibility of the existence of the present system is supported by the local minimum of the action. In addition, the stability of the system can be determined from the Wilson loop, and indicates that the present system, in the confining phase, is more stable than the Coulomb phase.
\par
In the center-of-mass frame, the fermion and antifermion are located at opposite positions with respect to the origin of $x$ axis ($y=z=0$). We treat here the localized fields $A^{a}_{{\rm (CL)}\mu}$, some of which are shifted from $A^{a}_{{\rm (CL)}\mu}=0$ to non-zero field for $1 \leq a \leq 3$, and expressed by
\begin{eqnarray}
A^{a}_{{\rm (CL)}t}(x)=\lambda^{a}P_{(0)}s_{\rm n}^{(0)}\exp(-|c^{(0)}_{\nu}x_{\nu}|),
\end{eqnarray}
\begin{eqnarray}
A^{a}_{{\rm (CL)}x}(x)=0,
\end{eqnarray}
\begin{eqnarray}
A^{a}_{{\rm (CL)}y}(x)=\lambda^{a}P_{(2)}s_{\rm n}^{(2)}\exp(-|c^{(2)}_{\nu}x_{\nu}|),
\end{eqnarray}
\begin{eqnarray}
A^{a}_{{\rm (CL)}z}(x)=\lambda^{a}P_{(3)}s_{\rm n}^{(3)}\exp(-|c^{(3)}_{\nu}x_{\nu}|),
\end{eqnarray}
where
\begin{eqnarray}
\nonumber
P_{(0)}=
(-\frac{1}{2}k^{-1/2}_{3}b^{(0)}_{\nu}x_{\nu})
\end{eqnarray}
\begin{eqnarray}
\times \frac{\exp(-|b^{\rm (M)}_{\mu\nu}x_{\mu}x_{\nu}|)}
{[1-\exp(-2|b^{\rm (M)}_{\mu\nu}x_{\mu}x_{\nu}|)]^{1/2}}.
\end{eqnarray}
For $a \geq 4$, the components of the field are given by $A^{a}_{{\rm (CL)}\mu}=0$. The quantities $\lambda^{a}$ are constants which depend on the superscript $a$ of the Lie algebra. In the center-of-mass frame, $b^{(0)}_{\nu}$ is $(0,a_{\rm c},0,0)$, and $c^{(0)}_{\nu}=c^{(2)}_{\nu}=c^{(3)}_{\nu}=(0,0,1/d,1/d)$, where $a_{\rm c}>0$ and $d>0$. The matrix mentioned above is represented as
\begin{eqnarray}
b^{\rm (M)}_{\mu\nu}=
\left[\begin{array}{cccc}
0                 &\frac{1}{2}a_{\rm c}&0           &0           \\
\frac{1}{2}a_{\rm c}&0                 &0           &0           \\
0                 &0                 &0         &0           \\
0                 &0                 &0    &0 
\end{array}\right].
\end{eqnarray}
The variable $s_{\rm n}^{(0)}$ is a product written as
\begin{eqnarray}
s_{\rm n}^{(0)}=
s_{{\rm n}t}^{(0)}s_{{\rm n}x}^{(0)}s_{{\rm n}y}^{(0)}s_{{\rm n}z}^{(0)},
\end{eqnarray}
and $s_{\rm n}^{(2)}$ as well as $s_{\rm n}^{(3)}$ have the corresponding form, where
\begin{eqnarray}
\hspace{-7ex}
s_{{\rm n}t}^{(0)}=s_{{\rm n}t}^{(2)}=s_{{\rm n}t}^{(3)}=+1
\hspace{5ex}
\mbox{ for all $t$},
\end{eqnarray}
\begin{eqnarray}
\hspace{-7ex}
s_{{\rm n}x}^{(0)}=s_{{\rm n}x}^{(2)}=s_{{\rm n}x}^{(3)}=+1
\hspace{5ex}
\mbox{ for all $x$},
\end{eqnarray}
\begin{eqnarray}
\hspace{-6ex}
s_{{\rm n}y}^{(0)}=s_{{\rm n}y}^{(2)}=s_{{\rm n}y}^{(3)}=+1
\hspace{5ex}
\mbox{ for $y \geq 0$},
\end{eqnarray}
\begin{eqnarray}
\hspace{-6ex}
s_{{\rm n}y}^{(0)}=-s_{{\rm n}y}^{(2)}=s_{{\rm n}y}^{(3)}=-1
\hspace{3ex}
\mbox{ for $y < 0$},
\end{eqnarray}
\begin{eqnarray}
\hspace{-5ex}
s_{{\rm n}z}^{(0)}=s_{{\rm n}z}^{(2)}=s_{{\rm n}z}^{(3)}=+1
\hspace{5ex}
\mbox{ for $z \geq 0$},
\end{eqnarray}
\begin{eqnarray}
\hspace{-5ex}
s_{{\rm n}z}^{(0)}=s_{{\rm n}z}^{(2)}=-s_{{\rm n}z}^{(3)}=-1
\hspace{3ex}
\mbox{ for $z < 0$}. 
\end{eqnarray}
\par
The field $A^{a}_{{\rm (CL)}t}$ for
$x \geq \epsilon_{x}$ 
is then denoted as
\begin{eqnarray}
A^{a}_{{\rm (CL)}t}(t,{\bf x})=
\lambda^{a}P_{(0)}(t,x)w_{t}(y)w_{t}(z),
\end{eqnarray}
where
\begin{eqnarray}
P_{(0)}(t,x)=\frac{1}{2}k^{-1/2}_{3}h(t,x),
\end{eqnarray}
and
\begin{eqnarray}
h(t,x)=  \frac{ - a_{\rm c} x \exp(-a_{\rm c}tx) }
{ [1-\exp(-2a_{\rm c}tx)]^{1/2} }
\hspace{2ex}
\mbox{ for $x \geq \epsilon_{x} $},
\end{eqnarray}
with the relation
\begin{eqnarray}
h(t,-x)=-h(t,x)
\hspace{2ex}
\mbox{ for $x < -\epsilon_{x} $}.
\end{eqnarray}
Here, $\epsilon_{x}$ is an infinitesimal positive quantity and we put $\epsilon_{x} \rightarrow 0$ after the calculations. The derivative is defined by $\partial_{x}h=\lim_{\Delta x \rightarrow +0}(\Delta h/(\Delta x)$ for $x \geq \epsilon_{x}$, and $\partial_{x}h=\lim_{\Delta x \rightarrow -0}(\Delta h/(\Delta x)$ for $x < -\epsilon_{x}$.
The function $w_{t}(y)$ is defined by
\begin{eqnarray}
w_{t}(y)=w(y)=
\left\{\begin{array}{ll}
+\exp(-\frac{y}{d}) & \mbox{ for $y \geq \epsilon_{y}$} \\
 &  \\
-\exp(+\frac{y}{d}) & \mbox{ for $y < -\epsilon_{y}$} \\
\end{array}\right. ,
\end{eqnarray}
with $\epsilon_{y}$ being an infinitesimal positive quantity putting $\epsilon_{y} \rightarrow 0$. The derivative is $\partial_{y}w=\lim_{\Delta y \rightarrow +0}(\Delta w/(\Delta y)$ for $y \geq \epsilon_{y}$, and $\partial_{y}w=\lim_{\Delta y \rightarrow -0}(\Delta w/(\Delta y)$ for $y < -\epsilon_{y}$.
The definitions for $z$ are the same as those for $y$.
\par
The fields $A^{a}_{{\rm (CL)}y}(t,{\bf x})$ and $A^{a}_{{\rm (CL)}z}(t,{\bf x})$ are
\begin{eqnarray}
A^{a}_{{\rm (CL)}y}(t,{\bf x})=
\lambda^{a}(\frac{d}{2})(\partial_{t}P_{(0)})w_{y}(y)w(z),
\end{eqnarray}
\begin{eqnarray}
A^{a}_{{\rm (CL)}z}(t,{\bf x})=
\lambda^{a}(\frac{d}{2})(\partial_{t}P_{(0)})w(y)w_{z}(z),
\end{eqnarray}
where
\begin{eqnarray}
w_{y}(y)=
\left\{\begin{array}{ll}
+w(y) & \mbox{ for $y \geq \epsilon_{y}$} \\
 &  \\
-w(y) & \mbox{ for $y < -\epsilon_{y}$} \\
\end{array}\right. ,
\end{eqnarray}
\begin{eqnarray}
w_{z}(z)=
\left\{\begin{array}{ll}
+w(z) & \mbox{ for $z \geq \epsilon_{z}$} \\
 &  \\
-w(z) & \mbox{ for $z < -\epsilon_{z}$} \\
\end{array}\right. .
\end{eqnarray}
Then, the localized functions of the vector potentials satisfy the Lorentz condition (derivatives are defined only in the region where the function is defined)
\begin{eqnarray}
\nonumber
\partial_{\mu}A^{a}_{{\rm (CL)}\mu}(x)
=\lambda^{a}
[(\partial_{t} P_{(0)})w(y)w(z)+0
\end{eqnarray}
\begin{eqnarray}
-\frac{1}{2}(\partial_{t} P_{(0)})w(y)w(z)
-\frac{1}{2}(\partial_{t} P_{(0)})w(y)w(z) ]=0.
\end{eqnarray}
\par
In the center-of-mass frame, we consider the region
$
0 < 
T_{\rm (b0)}
\leq \epsilon_{t}
\leq t \leq T_{\rm (b)}$,
$|x| \leq X_{\rm (b)}$, $|y| \leq Y_{\rm (b)}$ and $|z| \leq Z_{\rm (b)}$, and the field
$A^{a}_{\rm (CL)\mu}$
takes non-zero values in
the region contained within the above region under consideration,
where $\epsilon_{t}$ is the (renormalization) scale-invariant time as explained later.
The localized function of Yang-Mills fields here is localized in the region $|y| \leq d$ and $|z| \leq d$, where $d$ is the size (inverse decay constant) of the
$tx$
sheet and is taken to be $d \rightarrow 0$ after the calculations, reducing the localized field energy. (The region indicated by $|y| > d$ and $|z| > d$ is not important.) Note, even in this limit the dominant thin $tx$ sheet has a finite energy, and the Wilson loop can be calculated in the $tx$ plane.
\par
We next present the unlocalized function $A^{a}_{{\rm (CU)} \mu}$ which is a part of the classical vector potentials. Before we describe the unlocalized function, let us define $\rho^{a}_{t}$ which takes non-zero value for $t \geq
$
$ T_{\rm (b0)} $
as
\begin{eqnarray}
\rho^{a}_{t}(t,x,y,z)=
\left\{\begin{array}{l}
\lambda^{a}[
(\frac{1}{2}k_{3}^{-1/2})Q(t,x)\\
\mbox{ }\\
\hspace{2ex}
-\frac{2}{d^2}P_{(0)}(t,x)]w(y)w(z)\\
\mbox{ }\\
\hspace{6.5ex}
\mbox{ for $\epsilon_{x} \leq x \leq X_{c}$}\\
\mbox{ }\\
0
\hspace{5ex}
\mbox{ for $x > X_{c}$}\\
\end{array}\right. .
\end{eqnarray}
By writing $\rho^{a}_{t}(t,x,y,z)$ as $\rho^{a}_{t}(t,x,y,z,a_{c})$, we have
\begin{eqnarray}
\rho^{a}_{t}(t,x,y,z,a_{\rm c})=-\rho^{a}_{t}(t,x,y,z,-a_{\rm c})
\hspace{1ex}
\mbox{ for $x < -\epsilon_{x}$},
\end{eqnarray}
($\rho^{a}_{t}=0$ for $t < 
$
$ T_{\rm (b0)} $
for the corresponding derivative definition), where
\begin{eqnarray}
\nonumber
Q(t,x)=a_{\rm c}^{3}x^{3}Q_{1}(t,x) -2a_{\rm c}^{2}tQ_{2}(t,x)
+a_{\rm c}^{3}t^{2}xQ_{1}(t,x),
\end{eqnarray}
\begin{eqnarray}
\end{eqnarray}
\begin{eqnarray}
Q_{1}(t,x)=Q_{1/2}(t,x)+4Q_{3/2}(t,x)
+3Q_{5/2}(t,x), 
\end{eqnarray}
\begin{eqnarray}
Q_{2}(t,x)=Q_{1/2}(t,x)+Q_{3/2}(t,x), 
\end{eqnarray}
\begin{eqnarray}
Q_{1/2}(t,x)=\frac{\exp(-a_{\rm c}|tx|)}{[1-\exp(-2a_{\rm c}|tx|)]^{1/2}},
\end{eqnarray}
\begin{eqnarray}
Q_{3/2}(t,x)=\frac{\exp(-3a_{\rm c}|tx|)}{[1-\exp(-2a_{\rm c}|tx|)]^{3/2}},
\end{eqnarray}
\begin{eqnarray}
Q_{5/2}(t,x)=\frac{\exp(-5a_{\rm c}|tx|)}{[1-\exp(-2a_{\rm c}|tx|)]^{5/2}}.
\end{eqnarray}
Additionally, $\rho^{a}_{x}(t,{\bf x})$ is represented by
\begin{eqnarray}
\rho^{a}_{x}(t,{\bf x})=0.
\end{eqnarray}
\par
The unlocalized function is then given by
\begin{eqnarray}
\nonumber
A^{a}_{{\rm (CU)}t}(t,{\bf x})=
\int_{
T_{\rm (b0)}
}^{t} 
dt_{s} \int_{-X_{\rm (b)}}^{X_{\rm (b)}} dx_{s}
\int_{-Y_{\rm (b)}}^{Y_{\rm (b)}} dy_{s} \int_{-Z_{\rm (b)}}^{Z_{\rm (b)}} dz_{s}
\end{eqnarray}
\begin{eqnarray}
\times
A^{a}_{{\rm (CU)}t,s}(t,{\bf x};t_{s},{\bf x}_{s}),
\end{eqnarray}
where $A^{a}_{{\rm (CU)}t,s}(t,{\bf x};t_{s},{\bf x}_{s})$ with the subscript $^{\prime\prime} s ^{\prime\prime}$ is expressed in terms of the time-dependent four-dimensional Green's function in Euclidean spacetime by
\begin{eqnarray}
A^{a}_{{\rm (CU)}t,s}(t,{\bf x};t_{s},{\bf x}_{s})
=G_{4}(t,{\bf x};t_{s},{\bf x}_{s})\rho^{a}_{t}(t_{s},{\bf x}_{s}).
\end{eqnarray}
Here,
\begin{eqnarray}
\nonumber
G_{4}(t,{\bf x};t_{s},{\bf x}_{s})=S_{4}^{-1}
\end{eqnarray}
\begin{eqnarray}
\times
\frac{1}{(t-t_{s})^2+(x-x_{s})^2+(y-y_{s})^2+(z-z_{s})^2},
\end{eqnarray}
where
\begin{eqnarray}
S_{4}=\frac{(4-2)2\pi^{4/2}}{\Gamma(4/2)}=4\pi^{2},
\end{eqnarray}
with $\Gamma$ being the gamma function. In the integration of Green's function above, we replace $(t-t_{s})^2 + (x-x_{s})^2 +(y-y_{s})^2 +(z-z_{s})^2$ by $(t-t_{s})^2 + (x-x_{s})^2 +(y-y_{s})^2 +(z-z_{s})^2 +\epsilon$, where $\epsilon$ is an infinitesimal positive quantity with the limit $\epsilon \rightarrow 0$ taken after integration. Among integral points along the $y$ axis ($z$ axis), the point $y_{s}=0$ can be excluded based on the general mathematical definition of integral. (The integral regions with respect to $y$ are set to
$
-Y_{\rm (b)}
\leq y \leq -\epsilon_{y}$ and $+\epsilon_{y} \leq y \leq
Y_{\rm (b)}
$, with $\epsilon_{y}>0$, followed by $\epsilon_{y} \rightarrow 0$.)
\par
Since $w(-y_{s})=-w(y_{s})$; $w(-z_{s})=-w(z_{s})$ and $\rho^{a}_{t}$ is proportional to $w(y_{s})w(z_{s})$, we have
\begin{eqnarray}
\rho^{a}_{t}(t_{s},x_{s},-y_{s},z_{s})=-\rho^{a}_{t}(t_{s},x_{s},y_{s},z_{s}),
\end{eqnarray}
\begin{eqnarray}
\rho^{a}_{t}(t_{s},x_{s},y_{s},-z_{s})=-\rho^{a}_{t}(t_{s},x_{s},y_{s},z_{s}).
\end{eqnarray}
These relations lead to the cancellation of charge contributions to the Green's function integral at $y=0$ $(z=0)$, and we derive
\begin{eqnarray}
A^{a}_{{\rm (CU)}t}(t,{\bf x})=0
\hspace{2ex}
\mbox{ at $y=0$ $(z=0)$}.
\end{eqnarray}
\par
In contrast, the localized function $A^{a}_{{\rm (CL)}t}(t,x,y,z)$ has a finite value at the same point, $y=0$ $(z=0)$. Additionally, the localized function $A_{{\rm (CL)}t}(t,{\bf x})$ is proportional to $\exp(-a_{\rm c}|tx|)$, and is localized within the region where $a_{\rm c}|tx|$ is small, and $|y| \leq d$ and $|z| \leq d$. Conversely, the unlocalized function $A^{a}_{{\rm (CU)}t}(t,{\bf x})$ spreads even in the region where $a_{\rm c}|tx|$ is large, and $|y|>d$ and $|z|>d$. Therefore, the behavior of the unlocalized function is different from the localized function. Additionally,
\begin{eqnarray}
A^{a}_{{\rm (CU)}x}(t,{\bf x})=0,
\end{eqnarray}
because of $\rho^{a}_{x}(t_{s},x_{s},y_{s},z_{s})=0$.
\par
We can show that the classical vector potential $A^{a}_{\rm (C)\mu}(x)$, composed of the localized function $A^{a}_{{\rm (CL)}\mu}(x)$ and the unlocalized funct1ion $A^{a}_{{\rm (CU)}\mu}(x)$, satisfies the classical equations of motion. The field is
an
${\rm SU}(2)$ gauge field
embedded in ${\rm SU}(N)$
and denoted as $A^{a}_{\rm (C)\mu}(x)=\lambda^{a}\tilde{A}_{\rm (C)\mu}(x)$ for $1 \leq a \leq 3$ (where $\tilde{A}_{\rm (C)\mu}(x)$ is common for $1 \leq a \leq 3$ and $A^{a}_{\rm (C)\mu}(x)=0$ for $a \geq 4$ 
). 
In this case, by using the relation $f^{abc}=-f^{acb}$ and
\begin{eqnarray}
A^{b}_{\rm (C)\mu}A^{c}_{\rm (C)\nu}=
\lambda^{b}\lambda^{c}\tilde{A}_{\rm (C)\mu}\tilde{A}_{\rm (C)\nu}
=A^{c}_{\rm (C)\mu}A^{b}_{\rm (C)\nu},
\end{eqnarray}
which leads to $\sum_{b,c}g
_{\rm c}
f^{abc}A^{b}_{\rm (C)\mu}A^{c}_{\rm (C)\nu}=0$
($g_{\rm c}$ is the classical coupling constant),
we have the field tensor for $1 \leq a \leq 3$
\begin{eqnarray}
\nonumber
F^{a}_{\rm (C)\mu\nu}=\partial_{\mu}A^{a}_{\rm (C)\nu}
-\partial_{\nu}A^{a}_{\rm (C)\mu}-
\sum_{b,c}g
_{\rm c}
f^{abc}A^{b}_{\rm (C)\mu}A^{c}_{\rm (C)\nu}
\end{eqnarray}
\begin{eqnarray}
=\lambda^{a} \tilde{F}_{\rm (C)\mu\nu},
\end{eqnarray}
with
\begin{eqnarray}
\nonumber
\tilde{F}_{\rm (C)\mu\nu}=\partial_{\mu}\tilde{A}_{\rm (C)\nu}
-\partial_{\nu}\tilde{A}_{\rm (C)\mu}-
\sum_{b,c}g
_{\rm c}
f^{abc}\tilde{A}_{\rm (C)\mu}\tilde{A}_{\rm (C)\nu}
\end{eqnarray}
\begin{eqnarray}
=\partial_{\mu}\tilde{A}_{\rm (C)\nu}
-\partial_{\nu}\tilde{A}_{\rm (C)\mu},
\end{eqnarray}
and $F^{a}_{\rm (C)\mu\nu}=0$ for $a \geq 4$.
In addition, for the classical vector potential $A_{\rm (C)\mu}=\sum_{a}A^{a}_{\rm (C)\mu}T^{a}$, the Yang-Mills equations of
motion 
\begin{eqnarray}
\partial_{\mu}F_{\rm (C)\mu\nu}+ig
_{\rm c}
[A_{\rm (C)\mu},F_{\rm (C)\mu\nu}]=0,
\end{eqnarray}
with the relation
\begin{eqnarray}
\nonumber
[A_{\rm (C)\mu},F_{\rm (C)\mu\nu}]
=\sum_{b,c}A^{b}_{\rm (C)\mu}F^{c}_{\rm (C)\mu\nu}( T^{b}T^{c}-T^{c}T^{b})
\end{eqnarray}
\begin{eqnarray}
=\sum_{a,b,c}if^{bca}A^{b}_{\rm (C)\mu}F^{c}_{\rm (C)\mu\nu}T^{a},
\end{eqnarray}
can be rewritten as
\begin{eqnarray}
\partial_{\mu}F^{a}_{\rm (C)\mu\nu}
-\sum_{b,c}
g
_{\rm c}
f^{bca}A^{b}_{\rm (C)\mu}F^{c}_{\rm (C)\mu\nu}=0.
\end{eqnarray}
For the embedded field $A^{a}_{\rm (C)\mu}(x)=\lambda^{a}\tilde{A}_{\rm (C)\mu}(x)$, satisfying the relation $\sum_{b,c}g
_{\rm c}
f^{bca}A^{b}_{\rm (C)\mu}F^{c}_{\rm (C)\mu\nu}=0$,
we get the simplified Yang-Mills equations (due to the drop of the term with $\partial_{\mu}A^{a}_{(\rm C)\mu}$ in the present scheme)
\begin{eqnarray}
\nonumber
\partial_{\mu}F^{a}_{\rm (C)\mu\nu}=
\partial_{\mu}\partial_{\mu}A^{a}_{\rm (C)\nu}
-\partial_{\mu}\partial_{\nu}A^{a}_{(\rm C)\mu}
\end{eqnarray}
\begin{eqnarray}
=\partial^{2}_{\mu}A^{a}_{\rm (C)\nu}=
\partial^{2}_{\mu}(\lambda^{a}\tilde{A}_{\rm (C)\nu})=0.
\end{eqnarray}
\par
Using the relations given above, we derive that the total classical vector potential $A^{a}_{\rm (C) \mu}=A^{a}_{{\rm (CL)} \mu} + A^{a}_{{\rm (CU)} \mu}$ satisfies the simplified Yang-Mills equation for embedded fields as
\begin{eqnarray}
\partial^{2}_{\mu} A^{a}_{\rm (C) \nu}=\partial^{2}_{\mu}A^{a}_{{\rm (CL)} \nu}
+ \partial^{2}_{\mu} A^{a}_{{\rm (CU)} \nu}=0,
\end{eqnarray}
since
\begin{eqnarray}
\partial^{2}_{\mu}A^{a}_{{\rm (CL)}\nu}(t,{\bf x})
=-\rho^{a}_{\nu}(t,{\bf x}),
\end{eqnarray}
\begin{eqnarray}
\partial^{2}_{\mu}A^{a}_{{\rm (CU)}\nu}(t,{\bf x})
=\rho^{a}_{\nu}(t,{\bf x}).
\end{eqnarray}
(Since $\rho^{a}_{y}$ and $\rho^{a}_{z}$ are not important here, explicit expressions for them are not given. If required, these quantities can be obtained in the same way as $\rho^{a}_{t}$ and $\rho^{a}_{x}$.) Additionally, in the procedure to get $\partial^{2}_{\mu}A^{a}_{{\rm (CU)}\nu}(t,{\bf x})=\rho^{a}_{\nu}(t,{\bf x})$, we have used the following property of the Green's function
\begin{eqnarray}
\nonumber
\partial^{2}_{\mu}A^{a}_{{\rm (CU)}t}(t,{\bf x})
=
\end{eqnarray}
\begin{eqnarray}
\nonumber
\int_{T_{\rm (b0)}}^{t}
dt_{s} \int_{-X_{\rm (b)}}^{X_{\rm (b)}} dx_{s}
\int_{-Y_{\rm (b)}}^{Y_{\rm (b)}} dy_{s} \int_{-Z_{\rm (b)}}^{Z_{\rm (b)}} dz_{s}
\end{eqnarray}
\begin{eqnarray}
\nonumber
\times
\delta(t-t_{s})\delta(x-x_{s})\delta(y-y_{s})\delta(z-t_{s})
\rho^{a}_{t}(t_{s},{\bf x}_{s})
\end{eqnarray}
\begin{eqnarray}
=\rho^{a}_{t}(t,{\bf x}).
\end{eqnarray}
Thus, the existence of the present system is supported by the local minimum of the action. Next, we determine the stability of the system from the Wilson loop. The Wilson loop indicates that the present system is in the confining phase and is more stable than the Coulomb phase.
\par

\subsection{Classical Wilson loop}

For the classical vector potentials of the Yang-Mills fields $A^{a}_{\rm (C)\mu}(x)$ mentioned above, we here calculate the Euclidean Wilson loop. We show that the localized function $A^{a}_{{\rm (CL)}\mu}(x)$ leads to the confinement, while the unlocalized function $A^{a}_{{\rm (CU)}\mu}(x)$ does not contribute to the Wilson loop. Hence, we first calculate the Wilson loop for the localized function $A^{a}_{{\rm (CL)}\mu}(x)$, and then calculate the contribution from the unlocalized function $A^{a}_{{\rm (CU)}\mu}(x)$ to the Wilson loop. Because $A^{a}_{{\rm (CL)}\mu}(x)=0$
for
$a \geq 4$, the superscript will be restricted to $a \leq 3$ in the following relations. We consider a rectangle whose sides are parallel to the $x$ or $t$ axes, and let $x_{1}=-x_{2}$; $|x_{1}| \leq
X_{\rm (b)}
$;
$T_{\rm (b0)}$
$
\leq t_{1}
$
$<<$
$
t_{2} \leq T_{\rm (b)}$. Because $A^{a}_{{\rm (CL)}x}=0$, a line integral along $x$ axis for the Wilson loop is
\begin{eqnarray}
I^{a(1)}_{{\rm (CL)}}=\int_{x_{1}}^{x_{2}} dx
A^{a}_{{\rm (CL)}x} |_{t=t_{1}, y=0, z=0} = 0.
\end{eqnarray}
A line integral along $t$ axis becomes
\begin{eqnarray}
\nonumber
I^{a(2)}_{{\rm (CL)}}=\int_{t_{1}}^{t_{2}} dt
A^{a}_{{\rm (CL)}t} |_{x=x_{2}, y=0, z=0}
\end{eqnarray}
\begin{eqnarray}
=\lambda^{a}(\frac{1}{2}k^{-1/2}_{3})
[H_{{\rm (CL)}}(t_{2},x_{2})-H_{{\rm (CL)}}(t_{1},x_{2})],
\end{eqnarray}
where $H_{{\rm (CL)}}$ above is given by
\begin{eqnarray}
H_{{\rm (CL)}}(t,x)=\int dt^{\prime}
h(t^{\prime},x).
\end{eqnarray}
To integrate we change of the integration variable as $u=\exp({-a_{\rm c}xt})$ for $x \geq \epsilon_{x}$, and get
\begin{eqnarray}
\nonumber
H_{{\rm (CL)}}(t,x)= \int d u^{\prime} (1-u^{\prime 2})^{-1/2} =
-
\arccos(u)
\end{eqnarray}
\begin{eqnarray}
=
-
\arccos[\exp(-a_{\rm c}xt)],
\end{eqnarray}
as well as
\begin{eqnarray}
H_{{\rm (CL)}}(t,x)= -H_{{\rm (CL)}}(t,-x)
\hspace{2ex}
\mbox{ for $x < -\epsilon_{x}$ }.
\end{eqnarray}
Similarly, since $H_{{\rm (CL)}}(t,x)$ is an odd function with respect to $x$, using $x_{1}=-x_{2}$, we have
\begin{eqnarray}
I^{a(3)}_{{\rm (CL)}}=\int_{x_{2}}^{x_{1}} dx
A^{a}_{{\rm (CL)}x} |_{t=t_{2}, y=0, z=0} = 0,
\end{eqnarray}
\begin{eqnarray}
\nonumber
I^{a(4)}_{{\rm (CL)}}=\int_{t_{2}}^{t_{1}} dt
A^{a}_{{\rm (CL)}t} |_{x=x_{1}, y=0, z=0}
\end{eqnarray}
\begin{eqnarray}
\nonumber
=\lambda^{a}(\frac{1}{2}k^{-1/2}_{3})
[H_{{\rm (CL)}}(t_{1},x_{1})-H_{{\rm (CL)}}(t_{2},x_{1})]
\end{eqnarray}
\begin{eqnarray}
=\lambda^{a}(\frac{1}{2}k^{-1/2}_{3})
[H_{{\rm (CL)}}(t_{2},x_{2})-H_{{\rm (CL)}}(t_{1},x_{2})].
\end{eqnarray}
Summation of the results above yields
\begin{eqnarray}
\nonumber
I^{a}_{{\rm (CL)}}=I^{a(1)}_{{\rm (CL)}}+I^{a(2)}_{{\rm (CL)}}+I^{a(3)}_{{\rm (CL)}}
+I^{a(4)}_{{\rm (CL)}}
\end{eqnarray}
\begin{eqnarray}
\nonumber
=
-
\lambda^{a}(k^{-1/2}_{3})
[\arccos(\exp(-a_{\rm c}x_{2}t_{2}))
\end{eqnarray}
\begin{eqnarray}
-\arccos(\exp(-a_{\rm c}x_{2}t_{1})) ].
\end{eqnarray}
For small $a_{\rm c}x_{2}t_{1}$, the last term $\arccos(\exp(-a_{\rm c}x_{2}t_{1}))$ on the right-hand side above can be neglected, and this approximation does not significantly influence the  confinement behavior of the heavy fermion.
\par
Thus, by using $A^{a}_{{\rm (CL)}\mu}=0$ for $a \geq 4$, the classical Wilson loop becomes
\begin{eqnarray}
\nonumber
W_{\rm C}= {\rm Tr} [ \exp(-\sum_{a=1}^{3} ig
_{\rm c}
I^{a}_{{\rm (CL)}}T^{a}) ]
\end{eqnarray}
\begin{eqnarray}
\nonumber
={\rm Tr}\biggl \{ \exp[\sum_{a=1}^{3} ig
_{\rm c}
\lambda^{a}k^{-1/2}_{3}[\arccos(\exp(-a_{\rm c}x_{2}t_{2}))]T^{a}]
\biggr \}.
\end{eqnarray}
\begin{eqnarray}
\end{eqnarray}
The matrices $T^{a}$ have the properties
\begin{eqnarray}
{\rm Tr}( T^{a}T^{b} )=\frac{1}{2}\delta_{a,b},
\end{eqnarray}
\begin{eqnarray}
T^{a}T^{b}+T^{b}T^{a}=0 
\hspace{2ex}
\mbox{ for $a \neq b$},
\end{eqnarray}
(where $\delta_{a,b}$ is the Kronecker function) 
and the normalization constant $k_{3}$ is set to satisfy
\begin{eqnarray}
k^{-1}_{3}g
_{\rm c}
^2 
\sum_{a=1}^{3} (\lambda^{a})^{2}T^{a}T^{a}=I^{(3)},
\end{eqnarray}
(where $I^{(3)}_{i^{\prime}j^{\prime}}=\delta_{i^{\prime},j^{\prime}}$ for $i^{\prime}, j^{\prime}=1, 2)$.
We can then drop the traceless odd terms in
a power series
of the exponential function, and we have
\begin{eqnarray}
\nonumber
W_{\rm C}=[{\rm Tr}(I^{(3)})]
\biggl[\cos[\arccos(\exp(-a_{\rm c}x_{2}t_{2}))]\biggr]
\end{eqnarray}
\begin{eqnarray}
=\exp[-\frac{a_{\rm c}}{2}(x_{2}-x_{1})(t_{2} -t_{1})+\ln( {\rm Tr}(I^{(3)}) )],
\end{eqnarray}
with the help of $x_{1}=-x_{2}$ and
$
0 < T_{\rm (b0)} \leq
t_{1}<<t_{2}$.
Note that this area law is not obtained in the case of Abelian gauge fields, because the matrix $T^{a}$ does not appear for Abelian gauge fields. In addition, since ${\rm SU}(N)$ (such as ${\rm SU}(3)$) includes the ${\rm SU}(2)$ subgroup, and the condition that $A^{a}_{{\rm (CL)}\mu}=0$ for $a \geq 4$ is not imposed on quantum fluctuations, then, the derivation mentioned above works not only for ${\rm SU}(2)$ but also ${\rm SU}(N)$.
\par
We can show that the unlocalized function of the classical vector potential does not contribute to the Wilson loop. Since $A^{a}_{{\rm (CU)}x}(t,{\bf x})=0$, we have
\begin{eqnarray}
I^{a(1)}_{{\rm (CU)}}=\int_{x_{1}}^{x_{2}} dx
A^{a}_{{\rm (CU)}x}(t,{\bf x})|_{t=t_{1}, y=0, z=0}=0,
\end{eqnarray}
\begin{eqnarray}
I^{a(3)}_{{\rm (CU)}}=\int_{x_{2}}^{x_{1}} dx
A^{a}_{{\rm (CU)}x}(t,{\bf x})|_{t=t_{2}, y=0, z=0}=0.
\end{eqnarray}
The other contributions to the Wilson loop, from the integral parallel to the time axis, are
\begin{eqnarray}
I^{a(2)}_{{\rm (CU)}}=\int_{t_{1}}^{t_{2}} dt
A^{a}_{{\rm (CU)}t}(t,{\bf x})|_{x=x_{2}, y=0, z=0},
\end{eqnarray}
\begin{eqnarray}
I^{a(4)}_{{\rm (CU)}}=\int_{t_{2}}^{t_{1}} dt
A^{a}_{{\rm (CU)}t}(t,{\bf x})|_{x=x_{1}, y=0, z=0}.
\end{eqnarray}
As was shown while
demonstrating the difference between the classical localized function and the unlocalized function, at the integral points of the Wilson loop we have
\begin{eqnarray}
A^{a}_{{\rm (CU)}t}(t,{\bf x})=0
\hspace{2ex}
\mbox{ at $y=0$, $z=0$}.
\end{eqnarray}
Then,
\begin{eqnarray}
I^{a(2)}_{{\rm (CU)}}=I^{a(4)}_{{\rm (CU)}}=0,
\end{eqnarray}
and the contribution from the unlocalized function to the Wilson loop vanishes as
\begin{eqnarray}
I^{a}_{{\rm (CU)}}=I^{a(1)}_{{\rm (CU)}}+I^{a(2)}_{{\rm (CU)}}+I^{a(3)}_{{\rm (CU)}}
+I^{a(4)}_{{\rm (CU)}}=0.
\end{eqnarray}
Thus, the classical unlocalized function does not contribute to the Wilson loop.
\par

\subsection{String tension and the physical interpretation of confinement with classical vector potentials}

The aforementioned formalism has a parameter $a_{\rm c}$, which we now determine. There is no total charge density for this classical vector potentials of Yang-Mills fields composed of the localized function and the unlocalized function. The energy of the classical vector potentials at real time $t^{\prime}$ (that is used here) in Minkowski spacetime corresponding to Euclidean time $t$ is written by
\begin{eqnarray}
E_{\rm (C)}(t^{\prime})=E_{{\rm (CL)}}(t^{\prime})+E_{{\rm (CU)}}(t^{\prime}),
\end{eqnarray}
where $E_{{\rm (CL)}}(t^{\prime})$ and $E_{{\rm (CU)}}(t^{\prime})$ are the localized and unlocalized parts, respectively, and the energy associated with each has the same form
\begin{eqnarray}
E_{{\rm (CL)}}=\int dxdydz T_{00{\rm (CL)}},
\end{eqnarray}
where
\begin{eqnarray}
T_{00{\rm (CL)}}=  \frac{\partial L_{{\rm F}{\rm (CL)}}}
{\partial(\frac{\partial A_{{\rm (CL)}\nu}}{\partial x^{\prime}_{0}})}
\frac{\partial A_{{\rm (CL)}\nu}}{\partial x^{\prime}_{0}} -L_{{\rm F}{\rm (CL)}},
\end{eqnarray}
with
\begin{eqnarray}
L_{{\rm F}{\rm (CL)}}=-\frac{2}{4} {\rm Tr} (F_{{\rm (CL)}\mu\nu})^2.
\end{eqnarray}
The energy
of the unlocalized function is produced due to charges associated with the change of the localized function and the energy conservation owing to the classical equations of motion.
The classical vector potentials of the Yang-Mills field obey the classical equation of motion. 
\par
The parameter $a_{\rm c}$ in the localized function of the classical vector potential is determined from the
(renormalization) scale-invariant
energy
relation
at the
scale-invariant time
$\epsilon_{t^{\prime}}$,
an
time interval 
from the spacetime point, where the heavy fermion-antifermion pair is created.
The
energy of the heavy fermion-antiparticle pair in the confining potential decreases.
Taking into account that the classical
potential of the particle
exists within $|x| \leq \epsilon_{t^{\prime}}$ (setting the speed of light to unity, $c=1$), the decreasing energy of the particle and antiparticle in the confining linear potential is measured from the point $x=\epsilon_{t^{\prime}}+\epsilon^{\prime}_{x}$, with $\epsilon^{\prime}_{x}$ being an infinitesimal positive value, where the linear potential is zero. We then derive the
decrease in energy of the particle and antiparticle
via the linear potential
\begin{eqnarray}
E_{(\rm LP)}(\epsilon_{t^{\prime}})=|-\frac{a_{\rm c}}{2}(2\epsilon_{t^{\prime}})|.
\end{eqnarray}
\par
In the integral to obtain the energy of the localized function of the classical vector potential, quantities such as 
\begin{eqnarray}
\nonumber
[(\int_{-\infty}^{-\epsilon_{y}}dy+ \int_{\epsilon_{y}}^{+\infty}dy)
(\int_{-\infty}^{-\epsilon_{z}}dz+ \int_{\epsilon_{z}}^{+\infty}dz)
\end{eqnarray}
\begin{eqnarray}
\nonumber
\times [w(y)]^2[w(z)]^2]
\end{eqnarray}
\begin{eqnarray}
=4 \int_{\epsilon_{y}}^{\infty} \int_{\epsilon_{z}}^{\infty} dydz [w(y)]^2[w(z)]^2 \simeq  d^2,
\end{eqnarray}
which depend on the decay length $d$ of $w(y)$, vanish in the limit $d \rightarrow 0$. The dominant terms are the integral for squares of the field strength $[-\partial_{y} A^{a}_{{\rm (CL)}t^{\prime}}]^2$ in $[F^{a}_{{\rm (CL)}yt^{\prime}}]^2$ and $[-\partial_{z} A^{a}_{{\rm (CL)}t^{\prime}}]^2$ in $[F^{a}_{{\rm (CL)}zt^{\prime}}]^2$, which is independent of the decay length $d$. These yield
\begin{eqnarray}
\nonumber
E_{{\rm (CL)}}(\epsilon_{t^{\prime}})
=\{ 2 \int_{0}^{\infty} dx [\frac{1}{g
_{\rm c}
}h(\epsilon_{t^{\prime}},x)]^2 \}
\end{eqnarray}
\begin{eqnarray}
\nonumber
\times \frac{4}{2}\{ \int_{0}^{\infty} \int_{0}^{\infty} dydz 
[(\partial_{y} w(y)w(z))^2+
(w(y)\partial_{z}w(z))^2] \},
\end{eqnarray}
\begin{eqnarray}
\end{eqnarray}
(since the above integral does not depend on $d$, the limits $\epsilon_{y} \rightarrow 0$, $\epsilon_{z} \rightarrow 0$ and $\epsilon_{x} \rightarrow 0$ have been taken).
The
integral simplifies to
\begin{eqnarray}
\int_{0}^{\infty} dx
(h(\epsilon_{t^{\prime}},x))^2
=\zeta(3)\Gamma(3) a_{\rm c}^2\frac{1}{(2a_{\rm c}\epsilon_{t^{\prime}})^3},
\end{eqnarray}
where $\zeta(3)$ is a constant value of $\zeta$ function, and gives
\begin{eqnarray}
E_{{\rm (CL)}}(\epsilon_{t^{\prime}})
=\frac{\zeta(3)\Gamma(3)}{4 g
_{\rm c}
^2}\frac{1}{a_{\rm c}(\epsilon_{t^{\prime}})^3}.
\end{eqnarray}
The
scale-invariant energy
relation $E_{{\rm (CL)}}(\epsilon_{t^{\prime}})=E_{(\rm LP)}(\epsilon_{t^{\prime}})=|-a_{\rm c}\epsilon_{t^{\prime}}|$ (setting the speed of light to unity, $c=1$),
which states that the energy of the solitonlike confining field (object) is equal to the decrease in energy due to the linear potential,
results in
\begin{eqnarray}
a_{\rm c}=\frac{\zeta(3)^{1/2}\Gamma(3)^{1/2}}
{2g
_{\rm c}
\epsilon_{t^{\prime}}^2}.
\end{eqnarray}
The string tension, $\sigma$, then amounts to
\begin{eqnarray}
\sigma=\frac{1}{2}a_{\rm c}=\frac{\zeta(3)^{1/2}\Gamma(3)^{1/2}}
{4g
_{\rm c}
\epsilon_{t^{\prime}}^2}.
\end{eqnarray}
\par
The maximum of the particle-antiparticle distance is the light cone diameter 2$\epsilon_{t^{\prime}}$ at
the scale-invariant
time $\epsilon_{t^{\prime}}$ (measured from the pair creation spacetime point with $c=1$).
We set the maximum effective radius of the particle as $R_{\rm p}=\epsilon_{t^{\prime}}$, and then $D_{\rm p}=2R_{\rm p}$, where $D_{\rm p}$ is the maximum effective size (diameter) of the particle. Owing to the
(renormalization)
scale-invariant constant property of QCD, the continuum theory has a
scale-invariant energy, $\lambda_{\rm MOM}$. Since our classical solution is in the continuum scheme, the scale-invariant length $1 / \lambda_{\rm MOM}$ in the continuum theory is set to the maximum pair size, which is twice the maximum effective size of the particle, giving
\begin{eqnarray}
\frac{1}{\lambda_{\rm MOM}}=2D_{\rm p}=4R_{\rm p}=4\epsilon_{t^{\prime}}.
\end{eqnarray}
This implies that the maximum pair size is physically scale-invariant. Our string tension is then rewritten to
\begin{eqnarray}
\sigma=\frac{4\zeta(3)^{1/2}\Gamma(3)^{1/2}\lambda_{\rm MOM}^2}
{g
_{\rm c}
}.
\end{eqnarray}
\par
In the string tension analysis, the lattice gauge theory uses the relation \cite{AHas}
\begin{eqnarray}
\frac{\lambda_{\rm MOM}}{\lambda_{\rm Lat}}=83.5,
\end{eqnarray}
where $\lambda_{\rm Lat}$ is the scale-invariant energy in the lattice gauge theory. The original computed value for the square root of the string tension, $\sigma_{\rm Lat}$, of the lattice gauge theory was expressed as
\begin{eqnarray}
\frac{\sqrt{\sigma_{\rm Lat}}}{\lambda_{\rm Lat}}=167.
\end{eqnarray}
This computed ratio decreased \cite{HRot}, through
\begin{eqnarray}
\frac{\sqrt{\sigma_{\rm Lat}}}{\lambda_{\rm Lat}}=92,
\end{eqnarray}
to
\begin{eqnarray}
\frac{\sqrt{\sigma_{\rm Lat}}}{\lambda_{\rm Lat}}=77.
\end{eqnarray}
The above theoretical relations reproduces $\sqrt{\sigma_{\rm Lat}}=$420 MeV at $\lambda_{\rm MOM}=$210, 381 and 455 MeV, respectively. On the other hand, since our classical field is in the continuum scheme, the coupling constant, $g
_{\rm c}
$, is a known value, $(0.1184\times 4\pi)^{1/2}$ \cite{MPes}. Using $\Gamma(3)=2$ and $\zeta(3)=1.202$, our theoretical relation for the square root of the string tension reproduces $\sqrt{\sigma}=420$ MeV at $\lambda_{\rm MOM}=$186 MeV, which is compatible with the scale-invariant QCD energy of approximately 200 MeV.
Note, the integral of $[F^{a}_{{\rm (CL)}\mu\nu}]^2$ with respect to $x$ can also be confirmed numerically, if required. (Gaussian quadratures may be used if the strict integral is required \cite{Fuku82,StrSec}.)
\par
Now, we examine this fermion confinement state, expressed by the explicit function of the classical vector potentials. All $A^{a}_{{\rm (CL)}\mu}(t,x,y,z)$ is squeezed around the $x$ axis connecting the heavy fermion and antifermion, resulting in a linear potential. Note, the $A^{a}_{{\rm (CL)}t}(t,x,y,z)$ takes the opposite sign on inversion of coordinate $x$. This causes the Wilson loop a rotation in the $tx$ plane. Furthermore, the $A^{a}_{{\rm (CL)}t}(t,x,y,z)$ and related charges also take the opposite sign on inversion of coordinates $y$ and $z$. Owing to these properties, the classical local $A^{a}_{{\rm (CL)}t}(t,x,y,z)$ gives rise to the confinement, independent of the classical unlocalized $A^{a}_{{\rm (CU)}t}(t,x,y,z)$, whose contribution vanishes because of cancellation between terms with opposite signs.
\par
The confinement in our scheme is caused by a specific topological object with solitonlike properties corresponding to the localized function, which is a part of the classical vector potential satisfying the classical Yang-Mills equation. Since the localized function has the factor $\exp(-a_{\rm c}|tx|)$ which approaches to zero for large $|tx|$, the object expressed by the localized function is localized in spacetime. In addition, $H_{{\rm (CL)}}(t,x)$ derived by integration
(the integral for a part of the solution $h(t,x)$ in $A^a_{(\rm CL) t}(t,\bf x)$ of the classical Yang-Mills equation)
is related to the solution $\phi(t,x)$ of the following sine-Gordon equation \cite{PerSky}
\begin{eqnarray}
\partial^{2}_{\mu} \phi - \sin (\phi)=0.
\end{eqnarray}
The equation above has soliton solutions given by
\begin{eqnarray}
\phi_{-}(t,x)=4 \arctan [ \exp(-x) ],
\end{eqnarray}
\begin{eqnarray}
\phi_{+}(t,x)=4 \arctan [ \exp(+x) ].
\end{eqnarray}
The function $H_{{\rm (CL)}}(t,x)$ for $t=1$ and $a_{\rm c}=1$ is denoted in terms of the solution for the sine-Gordon equation \cite{PerSky} as
\begin{eqnarray}
H_{{\rm (CL)}}(1,x)
=-\arccos[ \tan \frac{1}{4}\phi_{-}(t,x)  ]
\hspace{2ex}
\mbox{ for $x \geq 0$},
\end{eqnarray}
\begin{eqnarray}
H_{{\rm (CL)}}(1,x)
=
\hspace{2ex}
\arccos[ \tan \frac{1}{4}\phi_{+}(t,x) ]
\hspace{2ex}
\mbox{ for $x < 0$}.
\end{eqnarray}
Thus we may say that the physical origin of the confinement is the specific solitonlike topological object, which disappears for large $|x|$. The solitonlike localized function cannot exist without the unlocalized function, since the localized function is necessary to satisfy the Yang-Mills equation.
\par
Additionally, the scale-invariant binding energy per fermion, which is the aforementioned decreasing energy of a fermion, is
\begin{eqnarray}
\nonumber
\epsilon_{\rm q}
=\frac{E_{\rm (LP)}(\epsilon_{t^{\prime}})} {2}
=\frac{\sigma (2 \epsilon_{t^{\prime}} )} {2}
\end{eqnarray}
\begin{eqnarray}
=\frac{\zeta(3)^{1/2}\Gamma(3)^{1/2}\lambda_{\rm MOM}}{g_{\rm c}}
\approx \lambda_{\rm MOM}.
\end{eqnarray}
This implies that the transition energy between the confinement phase and the deconfinement phase is approximately equal to $\lambda_{\rm MOM}$.
\par
Letting $\tau=1/k_{B}T$ with $k_{B}$ and $T$ being the Boltzmann constant and temperature, respectively, the periodic condition (along the temperature axis) imposed on the classical localized function is satisfied by the substitution of $t$ with $\tau^{\prime}$ and
\begin{eqnarray}
\nonumber
A^{a}_{{\rm (CL)}t}(t,x,y,z) \rightarrow A^{a}_{{\rm (CL)}\tau}(\tau^{\prime},x,y,z) 
\end{eqnarray}
\begin{eqnarray}
=A^{a}_{{\rm (CL)}t}(\tau^{\prime},x,y,z)+A^{a}_{{\rm (CL)}t}(\tau-\tau^{\prime},x,y,z).
\end{eqnarray}
Then, the one-way integral with respect to $\tau^{\prime}$ corresponding to the Wilson loop gives the Polyakov line \cite{Polya} as
\begin{eqnarray}
\nonumber
P_{\tau}=
{\rm Tr}\exp[-ig_{\rm c}
\sum_{a}(
\int_{\tau_{\epsilon}}^{\tau-\tau_{\epsilon}} d\tau^{\prime}
A^{a}_{{\rm (CL)}\tau} |_{x=x_{2}, y=0, z=0}
T^{a})]
\end{eqnarray}
\begin{eqnarray}
\nonumber
\approx
\cos[\arccos(\exp(-a_{\rm c}x_{2}\tau))
-\arccos(\exp(-a_{\rm c}x_{2}\tau_{\epsilon}))]
\end{eqnarray}
\begin{eqnarray}
\end{eqnarray}
where $\tau_{\epsilon}$ is a small positive quantity. 
By dropping the nonessential constant, the binding energy per fermion becomes
\begin{eqnarray}
\epsilon_{\rm q}(\tau)=-\ln P_{\tau}.
\end{eqnarray}
The Polyakov line, $P_{\tau}$, is smaller than unity for large $\tau$ at low temperatures, indicating the confinement phase, while $P_{\tau}$ approaches unity for small $\tau$ at high temperatures, indicating the deconfinement phase in some sense. The large change of $P_{\tau}$ occurs scale-invariantly around $\tau \approx 1/\lambda_{\rm MOM}$ at $x_{2}=1/\lambda_{\rm MOM}$.
\par

\section{Formulation for field expression in terms of real spacetime basis functions}

We present the expansion of fields in terms of
local
basis functions
in real spacetime,
required for the quantum fluctuation analysis presented in the next section. A four-dimensional hypercube with a central point $x_{\mu p}=x_{\mu (k,l,m,n)}$ (which we refer to as $(t_{k}, x_{l}, y_{m}, z_{n})$) is introduced, where the indices $k, l, m$ and $n$ in $p=(k,l,m,n)$ run from $1$ to $N_{(\mu)}$, respectively. The points $t_{k-1/2}$ and $t_{k+1/2}$ in the $t$-axis are defined as $t_{k-1/2}=t_{k}-(1/2)\Delta$ and $t_{k+1/2}=t_{k}+(1/2)\Delta$, where the infinitesimal positive $\Delta$ is given by $\Delta=t_{k+1}-t_{k}=x_{l+1}-x_{l}$, and we put $\Delta \rightarrow 0$ towards the construction of a continuum theory. (Corresponding quantities in the other axes are represented in the same way. As mentioned in our previous paper \cite{Fuku84}, the hypercube mentioned here can be transformed into a hyperhexahedron with curved sides, by mapping the hypercube in a parameter spacetime to the hyperhexahedron in real spacetime.) For a Euclidean spacetime restricted to the region $ T_{\rm (b0)} \leq t \leq T_{\rm (b)}$; $-X_{\rm (b)} \leq x \leq X_{\rm (b)}$; $-Y_{\rm (b)} \leq y \leq Y_{\rm (b)}$; $-Z_{\rm (b)} \leq z \leq Z_{\rm (b)}$, a set of basis functions around the central point of the hypercube is defined as
\begin{eqnarray}
\nonumber
\Omega^{4}_{p}(x)
=\Omega^{4}_{(k,l,m,n)}(t,{\bf x})
=\Omega^{E}_{k}(t)\Omega^{E}_{l}(x)\Omega^{E}_{m}(y)\Omega^{E}_{n}(z),
\end{eqnarray}
\begin{eqnarray}
\end{eqnarray}
\begin{eqnarray}
\Omega^{\delta 3}_{\mu p}(x)
=\Omega^{\delta 3-}_{\mu p}(x)-\Omega^{\delta 3+}_{\mu p}(x),
\end{eqnarray}
where $\Omega^{E}$ is represented (for example, as a function of $t$) by
\begin{eqnarray}
\Omega^{E}_{k}(t)=
\left\{\begin{array}{ll}
1 & \mbox{ for $t_{k-1/2} <  t < t_{k+1/2}$}\\
\mbox{ } & \mbox{ } \\
0 & \mbox{ for $t \leq t_{k-1/2}$ or
$t \geq t_{k+1/2}$}
\end{array}\right. ,
\end{eqnarray}
and $\Omega^{\delta 3-}_{0 p}$, $\Omega^{\delta 3+}_{0 p}$ and $\partial_{t}\Omega^{E}_{k}(t)|_{t=t_{k-1/2}}$ are defined as (for $\mu=0$)
\begin{eqnarray}
\nonumber
\Omega^{\delta 3-}_{0 p}(x)=
\Omega^{\delta 3-}_{0 (k,l,m,n)}(t,{\bf x})
\end{eqnarray}
\begin{eqnarray}
\nonumber
=\partial_{t}\Omega^{E}_{k}(t)|_{t=t_{k-1/2}}
\Omega^{E}_{l}(x)\Omega^{E}_{m}(y)\Omega^{E}_{n}(z)\Delta
\end{eqnarray}
\begin{eqnarray}
=\delta (t - t_{k-1/2})
\Omega^{E}_{l}(x)\Omega^{E}_{m}(y)\Omega^{E}_{n}(z)\Delta,
\end{eqnarray}
\begin{eqnarray}
\nonumber
\Omega^{\delta 3+}_{0 p}(x)
=-\partial_{t}\Omega^{E}_{k}(t)|_{t=t_{k+1/2}}
\Omega^{E}_{l}(x)\Omega^{E}_{m}(y)\Omega^{E}_{n}(z)\Delta
\end{eqnarray}
\begin{eqnarray}
=\delta (t - t_{k+1/2})
\Omega^{E}_{l}(x)\Omega^{E}_{m}(y)\Omega^{E}_{n}(z)\Delta.
\end{eqnarray}
In the definition of $\Omega^{\delta 3}_{\mu p}$ above, the derivative is multiplied by $\Delta$ in order to make the dimension for the latter basis function $\Omega^{\delta 3}_{\mu p}(x)$ equal to that for the former basis function $\Omega^{4}_{p}(x)$.
\par
The fields are now described, in terms of $\Omega^{4}_{q}(x)$ and
$\Omega^{\delta 3}_{\mu p}(x)$, as
\begin{eqnarray}
A^{a}_{\mu}(x)=\sum_{p}[A^{a}_{(P1)\mu p}\Omega^{4}_{p}(x)
+A^{a}_{(P2) p}\Omega^{\delta 3}_{\mu p}(x)].
\end{eqnarray}
Additionally, the Dirac fields are denoted as
\begin{eqnarray}
\psi(x)=\sum_{p}
\psi_{p}\Omega^{4}_{p}(x)
\exp [i \alpha^{0}(x)+ i\sum_{a}\alpha^{a}(x) T^{a}],
\end{eqnarray}
with
\begin{eqnarray}
\alpha^{a}(x)=\sum_{p}\alpha^{a}_{p}\Omega^{4}_{p}(x)
=\sum_{p}\alpha^{\prime a}_{p}\Omega^{4}_{p}(x)\Delta.
\end{eqnarray}
In the case of non-Abelian gauge fields, the conventional gauge transformation for local fields is
\begin{eqnarray}
\delta A^{a}_{\mu}(x)=-\partial_{\mu} \delta\alpha^{a}(x)
-\sum_{b,c} f^{bca}\delta\alpha^{b}(x)A^{c}_{\mu}(x).
\end{eqnarray}
In the present formalism, the term corresponding to the first term in the conventional gauge transformation, has the form
\begin{eqnarray}
\nonumber
-\partial_{\mu} \delta\alpha^{a}(x)
=-\partial_{\mu} \sum_{p} \delta\alpha^{\prime a}_{p}
\Omega^{4}_{p}(x)\Delta
=-\sum_{p}\delta\alpha^{\prime a}_{p}
\Omega^{\delta 3}_{\mu p}(x).
\end{eqnarray}
\begin{eqnarray}
\end{eqnarray}
\par
By using the relations
\begin{eqnarray}
\Omega^{4}_{p}(x)\Omega^{4}_{q}(x)=\delta_{p,q}\Omega^{4}_{p}(x),
\end{eqnarray}
\begin{eqnarray}
\Omega^{4}_{p}(x)\Omega^{\delta 3}_{\mu q}(x)=0,
\end{eqnarray}
the term in the present formalism, corresponding to the second term in the conventional gauge transformation, can be expressed in terms of the basis functions
\begin{eqnarray}
\nonumber
\delta\alpha^{b}(x)A^{c}_{\mu}(x)
=[\sum_{p}\delta\alpha^{b}_{p}\Omega^{4}_{p}(x)]
\end{eqnarray}
\begin{eqnarray}
\nonumber
\times
[\sum_{q}(A^{c}_{(P1) \mu q}\Omega^{4}_{q} (x)
+A^{c}_{(P2)q}\Omega^{\delta 3}_{\mu q}(x) )]
\end{eqnarray}
\begin{eqnarray}
=\sum_{p}\delta\alpha^{b}_{p}A^{c}_{(P1) \mu p}\Omega^{4}_{p} (x).
\end{eqnarray}
Therefore, the total gauge transformation in the present formalism is given by
\begin{eqnarray}
\nonumber
\delta A^{a}_{\mu}(x)
=-\sum_{p} \delta\alpha^{\prime a}_{p}
\Omega^{\delta 3}_{\mu p}(x)
\end{eqnarray}
\begin{eqnarray}
-\sum_{b,c}f^{bca}\sum_{p}
\delta\alpha^{b}_{p}A^{c}_{(P1) \mu p}\Omega^{4}_{p} (x).
\end{eqnarray}
By equating the above relation to
\begin{eqnarray}
\nonumber
\delta A^{a}_{\mu}(x)=
\sum_{p}\delta A^{a}_{(P1)\mu p}\Omega^{4}_{p}(x)
+\sum_{p}\delta A^{a}_{(P2) p}
\Omega^{\delta 3}_{\mu p}(x),
\end{eqnarray}
\begin{eqnarray}
\end{eqnarray}
we get following relations for gauge invariance:
\begin{eqnarray}
\delta A^{a}_{(P1)\mu p}=-\sum_{b,c}f^{bca}
\delta\alpha^{b}_{p}A^{c}_{(P1) \mu p},
\end{eqnarray}
\begin{eqnarray}
\delta A^{a}_{(P2) p}=-\delta\alpha^{\prime a}_{p}.
\end{eqnarray}
\par
The aforementioned step function takes zero value at the upper and lower boundaries of the region, between which the value of the step function is unity. The advanced basis set of step functions is defined so as to contain step functions, taking unity at the above upper and lower boundaries as
\begin{eqnarray}
\Omega^{E}_{k^{\prime\prime}}(t)=
\left\{\begin{array}{ll}
1 & \mbox{ for $t_{k^{\prime\prime}-1/2} \le  t \le t_{k^{\prime\prime}+1/2}$}\\
\mbox{ } & \mbox{ } \\
0 & \mbox{ for $t < t_{k^{\prime\prime}-1/2}$ or
$t > t_{k^{\prime\prime}+1/2}$}
\end{array}\right. ,
\end{eqnarray}
that are inserted between the original step functions without influencing the results. The site of the above basis function is indicated with double prime. The step function has non-vanishing cross terms with the derivative as
\begin{eqnarray}
\Omega^{4}_{p^{\prime\prime}}(x)\Omega^{\delta 3}_{\mu q}(x)=\delta_{p^{\prime\prime},q}\Omega^{\delta 3}_{\mu q}(x).
\end{eqnarray}
This leads to the relation
\begin{eqnarray}
\delta A^{a}_{(P2) p^{\prime\prime}}=-\delta\alpha^{\prime a}_{p^{\prime\prime}}
-\sum_{b,c}f^{bca}
\delta\alpha^{a}_{p^{\prime\prime}}A^{c}_{(P2) \mu p^{\prime\prime}},
\end{eqnarray}
for the gauge invariance.

\section{Quantum fluctuations}

\subsection{Action for quantum fluctuations around the classical field}

The field $A^{a}_{\mu}(x)$ in the present scheme consists of the classical field $A^{a}_{\rm (C)\mu}(x)$ and the quantum fluctuations $A^{a}_{{\rm (Q)}\mu}(x)$ around this classical field. The quantum fluctuation is for a ${\rm SU}(N)$ gauge field (such as ${\rm SU}(3)$) and $A^{a}_{{\rm (Q)}\mu}(x)$ may take non-zero value not only for $1 \leq a \leq 3$ but also for $a \geq 4$ unlike the classical field. (This property may also work for other fields with the same/generalized or similar algebra structure beyond ${\rm SU}(N)$.) Since the present scheme is gauge invariant, we use the gauge transformation to cancel the coefficients of the basis set of $\Omega^{3}_{p}(x)$ to give the following representation for quantum fluctuations
\begin{eqnarray}
A^{a}_{{\rm (Q)} \mu}(x)=\sum_{p} A^{a}_{({\rm Q}) \mu p} \Omega^{4}_{p}(x).
\end{eqnarray}
(Although the classical field is not expressed in terms of basis functions in this paper, such an expression may be possible.)
In the action, the first-derivative terms for quantum fluctuations around the classical field vanish due to the classical equations of motion.
\par
\par
The terms of the Yang-Mills field to be considered are
\begin{eqnarray}
S=S^{(2)}+S^{(3)}+S^{(4)},
\end{eqnarray}
where
\begin{eqnarray}
S^{(2)}=\frac{1}{2}\sum_{a} \int d^4x 
(\partial_{\nu} A^{a}_{{\rm (Q)} \mu}\partial_{\nu}A^{a}_{{\rm (Q)} \mu}),
\end{eqnarray}
\begin{eqnarray}
\nonumber
S^{(3)}=-g \sum_{a,b,c} f^{abc} \int d^4x
(A^{b}_{{\rm (Q)} \mu}A^{c}_{{\rm (Q)} \nu}\partial_{\mu}A^{a}_{\rm (C) \nu}
\end{eqnarray}
\begin{eqnarray}
\nonumber
+A^{b}_{{\rm (Q)} \mu}A^{c}_{\rm (C) \nu}\partial_{\mu}A^{a}_{{\rm (Q)} \nu}
+A^{b}_{\rm (C) \mu}A^{c}_{{\rm (Q)} \nu}\partial_{\mu}A^{a}_{{\rm (Q)} \nu}
\end{eqnarray}
\begin{eqnarray}
+A^{b}_{\rm (Q) \mu}A^{c}_{{\rm (Q)} \nu}\partial_{\mu}A^{a}_{{\rm (Q)} \nu}),
\end{eqnarray}
\begin{eqnarray}
\nonumber
S^{(4)}=
\end{eqnarray}
\begin{eqnarray}
\nonumber
\frac{g^{2}}{4} \sum_{a,b,c,d,e} f^{abc}f^{ade} \int d^4x
(A^{b}_{{\rm (Q)} \mu} A^{c}_{{\rm (Q)} \nu} A^{d}_{\rm (C) \mu} A^{e}_{\rm (C) \nu}
\end{eqnarray}
\begin{eqnarray}
\nonumber
+A^{b}_{{\rm (Q)} \mu} A^{c}_{\rm (C) \nu} A^{d}_{{\rm (Q)} \mu} A^{e}_{\rm (C) \nu}
+A^{b}_{{\rm (Q)} \mu} A^{c}_{\rm (C) \nu} A^{d}_{\rm (C) \mu} A^{e}_{{\rm (Q)} \nu}
\end{eqnarray}
\begin{eqnarray}
\nonumber
+A^{b}_{\rm (C) \mu} A^{c}_{{\rm (Q)} \nu} A^{d}_{{\rm (Q)} \mu} A^{e}_{\rm (C) \nu}
+A^{b}_{\rm (C) \mu} A^{c}_{{\rm (Q)} \nu} A^{d}_{\rm (C) \mu} A^{e}_{{\rm (Q)} \nu}
\end{eqnarray}
\begin{eqnarray}
\nonumber
+A^{b}_{\rm (C) \mu} A^{c}_{\rm (C) \nu} A^{d}_{{\rm (Q)} \mu} A^{e}_{{\rm (Q)} \nu}
+A^{b}_{\rm (Q) \mu} A^{c}_{\rm (Q) \nu} A^{d}_{{\rm (Q)} \mu} A^{e}_{{\rm (C)} \nu}
\end{eqnarray}
\begin{eqnarray}
\nonumber
+A^{b}_{\rm (Q) \mu} A^{c}_{\rm (Q) \nu} A^{d}_{{\rm (C)} \mu} A^{e}_{{\rm (Q)} \nu}
+A^{b}_{\rm (Q) \mu} A^{c}_{\rm (C) \nu} A^{d}_{{\rm (Q)} \mu} A^{e}_{{\rm (Q)} \nu}
\end{eqnarray}
\begin{eqnarray}
\nonumber
+A^{b}_{\rm (C) \mu} A^{c}_{\rm (Q) \nu} A^{d}_{{\rm (Q)} \mu} A^{e}_{{\rm (Q)} \nu}
+A^{b}_{\rm (Q) \mu} A^{c}_{\rm (Q) \nu} A^{d}_{{\rm (Q)} \mu} A^{e}_{{\rm (Q)} \nu}).
\end{eqnarray}
\begin{eqnarray}
\end{eqnarray}
\par
The cubic term of the above action is small, because the quantum coupling constant is weak for the small lattice spacing; the quartic term is smaller than the cubic term. This leaves only the quadratic term. The above statement is further supported by the following relations: the action in the Feynman gauge is positive as
\begin{eqnarray}
S=\sum_{a} \int d^4x 
[\frac{1}{4}F^{a}_{\mu\nu}F^{a}_{\mu\nu}
+\frac{1}{2}\partial_{\mu} A^{a}_{\mu}\partial_{\mu}A^{a}_{\mu}].
\end{eqnarray}
By writing the path integral in the form
\begin{eqnarray}
I_{A}=\frac{1}{Z_{{\rm N}A}} \int D[A^{a}_{\mu}]\exp(-S)=C_{A},
\end{eqnarray}
with a normalization constant $Z_{{\rm N}A}$, the integral over the region $\sum_{a\mu}(A^{a}_{\mu})^2 > B_{A}$ for a positive value $B_{A}$ amounts to a value less than a small positive value $\epsilon_{{\rm B}A}$. The path integral over the region $\sum_{a\mu}(A^{a}_{\mu})^2 \leq B_{A}$ truncating large fluctuations has a finite value multiplied by the coupling constant.
\par
The term $A^{b}_{{\rm (Q)} \mu} A^{c}_{\rm (C) \nu} A^{d}_{{\rm (Q)} \mu} A^{e}_{\rm (C) \nu}$ for $b=d$
gives
rise to a local
square
mass (energy gap) for quantum fluctuations of the Yang-Mills field.
In the action
\begin{eqnarray}
\nonumber
S^{(41)}_{\rm (CQ)}=
\frac{1}{4}g^2f^{abc}f^{ade} \int d^4x
A^{b}_{{\rm (Q)} \mu} A^{c}_{{\rm (C)} \nu} A^{d}_{\rm (Q) \mu} A^{e}_{\rm (C) \nu},
\end{eqnarray}
\begin{eqnarray}
\end{eqnarray}
the remaining component for the case of small $d$ in $w(y_s)=\exp(-y_s/d)$, which means that the solitonlike object of the classical localized function is thin, is
\begin{eqnarray}
\nonumber
S^{(41)}_{\rm (CQ)}=
\frac{1}{4}g^2f^{abc}f^{abe} \int d^4x
A^{b}_{{\rm (Q)} \mu} A^{c}_{{\rm (C)} t} A^{e}_{{\rm (C)} t} A^{b}_{{\rm (Q)} \mu}.
\end{eqnarray}
\begin{eqnarray}
\end{eqnarray}
For this case, the vector potential includes Green's function and the charge density as
\begin{eqnarray}
\nonumber
A^{a}_{{\rm (C)}t}(t,{\bf x}) \approx A^{a}_{{\rm (CU)}t}(t,{\bf x})
\end{eqnarray}
\begin{eqnarray}
\nonumber
=\int_{
T_{\rm (b0)}
}^{t} 
dt_{s} \int_{-X_{\rm (b)}}^{X_{\rm (b)}} dx_{s}
\int_{-Y_{\rm (b)}}^{Y_{\rm (b)}} dy_{s} \int_{-Z_{\rm (b)}}^{Z_{\rm (b)}} dz_{s}
\end{eqnarray}
\begin{eqnarray}
\nonumber
\times
\frac{\rho^{a}_{t}(t_{s},x_{s},y_{s},z_{s})}{4\pi^2[(t-t_{s})^2+(x-x_{s})^2+(y-y_{s})^2+(z-z_{s})^2]},
\end{eqnarray}
\begin{eqnarray}
\end{eqnarray}
and the charge density contains the solitonlike function as
\begin{eqnarray}
\nonumber
\rho^{a}_{t}(t_{s},x_{s},y_{s},z_{s}) \approx
\end{eqnarray}
\begin{eqnarray}
-\frac{1}{2d^2} \frac{ - a_{\rm c} x_s \exp(-a_{\rm c}t_s x_s) }
{ [1-\exp(-2a_{\rm c}t_s x_s)]^{1/2} }
w(y_s)w(z_s).
\end{eqnarray}
The approximate integral of the vector potential,
\begin{eqnarray}
\nonumber
A^{c}_{{\rm (C)} t} \approx \int d^4x_s
[\frac{1}{d}\exp(-\frac{y_s}{d})]
[\frac{1}{d}\exp(-\frac{z_s}{d})]
\frac{1}{r_{s}^2}
\end{eqnarray}
\begin{eqnarray}
\times
\frac{ - a_{\rm c} x_s \exp(-a_{\rm c}t_s x_s) }
{ [1-\exp(-2a_{\rm c}t_s x_s)]^{1/2} },
\end{eqnarray}
with $r_{s}^2=(t-t_{s})^2+(x-x_{s})^2+(y-y_{s})^2+(z-z_{s})^2]$, has
\begin{eqnarray}
\nonumber
\frac{1}{d} \int_{0}^{+\infty} dy_s \exp(-\frac{y_s}{d})
=\frac{1}{d} \int_{0}^{+\infty} dz_s \exp(-\frac{z_s}{d})
\end{eqnarray}
\begin{eqnarray}
\nonumber
=\frac{1}{d}(\int_{0}^{y} dy_s + \int_{y}^{+\infty} dy_s)
\exp[\frac{(-(y_s-y)-y)}{d}]=1,
\end{eqnarray}
\begin{eqnarray}
\end{eqnarray}
and
\begin{eqnarray}
\int dt_s \frac{ - a_{\rm c} x_s \exp(-a_{\rm c}t_s x_s) }
{ [1-\exp(-2a_{\rm c}t_s x_s)]^{1/2} } \approx 1.
\end{eqnarray}
The integral with respect to $x_s$ for the pair size,
\begin{eqnarray}
A^{c}_{{\rm (C)} t} \approx \int dx_s \frac{1}{r_s^2},
\end{eqnarray}
produces the scale-invariant relation and the renormalizability, including the charge coupling, corresponding to the scale-invariant energy as
\begin{eqnarray}
m_{\rm YM} \approx \lambda_{\rm MOM},
\end{eqnarray}
and $m_{\rm YM}^2 \approx \lambda_{\rm MOM}^2$, where $m_{\rm YM}$ is the Yang-Mills local mass.
\par
The first term we consider is
\begin{eqnarray}
\nonumber
S^{(2)a}_{\rm Q,Q}=\frac{1}{2}\int d^4x 
\partial_{\nu} A^{a}_{{\rm (Q)} \mu}\partial_{\nu}A^{a}_{{\rm (Q)} \mu}
\end{eqnarray}
\begin{eqnarray}
=\frac{1}{2}\int d^4x 
(\partial_{\nu}\sum_{p}A^{a}_{({\rm Q}) \mu p}\Omega^{4}_{p})
(\partial_{\nu}\sum_{q}A^{a}_{({\rm Q}) \mu q}\Omega^{4}_{q}),
\end{eqnarray}
which is rewritten using
\begin{eqnarray}
\partial_{\nu}\Omega^{4}_{p}
=\frac{1}{\Delta}\Omega^{\delta 3-}_{\nu p}-
\frac{1}{\Delta}\Omega^{\delta 3+}_{\nu p},
\end{eqnarray}
as
\begin{eqnarray}
\nonumber
S^{(2)a}_{\rm Q,Q}=S^{(2)a}_{\rm Q-,Q-}-S^{(2)a}_{\rm Q-,Q+}
-S^{(2)a}_{\rm Q+,Q-}+S^{(2)a}_{\rm Q+,Q+},
\end{eqnarray}
\begin{eqnarray}
\end{eqnarray}
where the former subscript ${\rm Q}\pm$ and the latter ${\rm Q}\pm$ in $S^{(2)a}_{{\rm Q}\pm ,{\rm Q}\pm}$ refer to the term including the former $\Omega^{\delta 3\pm}_{\nu p}$ and the latter $\Omega^{\delta 3\pm}_{\nu q}$ like
\begin{eqnarray}
\nonumber
S^{(2)a}_{{\rm Q}-,{\rm Q}-}
=\frac{1}{2}(\sum_{p}\sum_{q}
A^{a}_{({\rm Q}) \mu p}A^{a}_{({\rm Q}) \mu q}
\end{eqnarray}
\begin{eqnarray}
\times
\int d^4x
\frac{1}{\Delta}\Omega^{\delta 3-}_{\nu p}
\frac{1}{\Delta}\Omega^{\delta 3-}_{\nu q}).
\end{eqnarray}
\par
For $x_{\nu}=t$, we get
\begin{eqnarray}
\nonumber
S^{(2)a}_{{\rm Q}-,{\rm Q}-,t}=
\end{eqnarray}
\begin{eqnarray}
\nonumber
\frac{1}{2}[\sum_{k,l,m,n} \sum_{K,L,M,N}
A^{a}_{({\rm Q}) \mu (k,l,m,n)}A^{a}_{({\rm Q}) \mu (K,L,M,N)}
\end{eqnarray}
\begin{eqnarray}
\nonumber
\times
\int d^4x \delta(t-t_{k-1/2})
\Omega^{E}_{l}(x)\Omega^{E}_{m}(y)\Omega^{E}_{n}(z)
\end{eqnarray}
\begin{eqnarray}
\times
\delta(t-t_{K-1/2})
\Omega^{E}_{L}(x)\Omega^{E}_{M}(y)\Omega^{E}_{N}(z)].
\end{eqnarray}
The basis functions above have the following properties
\begin{eqnarray}
\int dx  \Omega^{E}_{l}(x)\Omega^{E}_{L}(x)
=\delta_{l,L}\Delta,
\end{eqnarray}
\begin{eqnarray}
\nonumber
\sum_{K} \int dt G_{k}(K) \delta(t-t_{k-1/2})
\delta(t-t_{K-1/2})
\end{eqnarray}
\begin{eqnarray}
\nonumber
=\sum_{K} G_{k}(K) \delta(t_{K-1/2}-t_{k-1/2})
\end{eqnarray}
\begin{eqnarray}
=\int dt^{\prime} \frac{1}{\Delta}G_{k}(K)
\delta(t^{\prime}-t_{k-1/2}),
\end{eqnarray}
where $t^{\prime}=t_{K-1/2}=(K-1/2)\Delta + t_{0}$ with constant $t_{0}$, and $G_{k}(K)=G_{k,K}$. The $\delta$ function above makes sense as an integral and results in a finite value for $t_{K-1/2}=t_{k-1/2}$, that is, for $k=K$. Then, the above relation becomes
\begin{eqnarray}
\nonumber
\frac{1}{\Delta}G_{k}(K)\delta_{(K-1/2),(k-1/2)}
=\frac{1}{\Delta}G_{k}(K)\delta_{k,K}
\end{eqnarray}
\begin{eqnarray}
=\sum_{K}\frac{1}{\Delta} G_{k,K} \delta_{k,K}.
\end{eqnarray}
With the use of the relations above, we derive
\begin{eqnarray}
\nonumber
S^{(2)a}_{Q-,Q-}=
\end{eqnarray}
\begin{eqnarray}
\nonumber
\frac{1}{2}(\sum_{k,l,m,n} \sum_{K,L,M,N}
A^{a}_{({\rm Q}) \mu (k,l,m,n)} A^{a}_{({\rm Q}) \mu (K,L,M,N)}
\end{eqnarray}
\begin{eqnarray}
\times
\frac{4}{\Delta}\Delta\Delta\Delta
\delta_{k,K}\delta_{l,L}\delta_{m,M}\delta_{n,N}).
\end{eqnarray}
Roughly speaking, this corresponds to the relation of the symbolic form $\int dtdxdydz \partial_{t}\partial_{t} \propto (\Delta)^{4}(\Delta^{-2})$.
We do not consider
the cubic or quartic term as aforementioned.
\par
Now, the eigenvalues of the matrix which describes the quadratic term $S^{(2)a}_{{\rm Q},{\rm Q}}$ are considered. The matrix to be diagonalized has some analogy with the matrix in a vibrational problem \cite{SlaFra}. We express $S^{(2)a}_{{\rm Q},{\rm Q}}$ by
\begin{eqnarray}
S^{(2)a}_{{\rm Q},{\rm Q}}
=\sum_{p,q}A^{a}_{({\rm Q})\mu p}M^{{\rm Q},{\rm Q}}_{pq} A^{a}_{({\rm Q})\mu q},
\end{eqnarray}
which are represented using the matrix elements by
\begin{eqnarray}
\nonumber
S^{(2)a}_{{\rm Q},{\rm Q}}=
\end{eqnarray}
\begin{eqnarray}
\nonumber
\frac{1}{2} \sum_{k,l,m,n}\sum_{K,L,M,N}
[A^{a}_{({\rm Q}) \mu (k,l,m,n)} A^{a}_{({\rm Q}) \mu (K,L,M,N)}
\end{eqnarray}
\begin{eqnarray}
\nonumber
\times (
2\delta_{K,k}\delta_{L,l}\delta_{M,m}\delta_{N,n}
-\delta_{K,k-1}\delta_{L,l}\delta_{M,m}\delta_{N,n}
\end{eqnarray}
\begin{eqnarray}
\nonumber
-\delta_{K,k+1}\delta_{L,l}\delta_{M,m}\delta_{N,n}
+2\delta_{K,k}\delta_{L,l}\delta_{M,m}\delta_{N,n}
\end{eqnarray}
\begin{eqnarray}
\nonumber
 -\delta_{K,k}\delta_{L,l-1}\delta_{M,m}\delta_{N,n}
 -\delta_{K,k}\delta_{L,l+1}\delta_{M,m}\delta_{N,n}
\end{eqnarray}
\begin{eqnarray}
\nonumber
+2\delta_{K,k}\delta_{L,l}\delta_{M,m}\delta_{N,n}
 -\delta_{K,k}\delta_{L,l}\delta_{M,m-1}\delta_{N,n}
\end{eqnarray}
\begin{eqnarray}
\nonumber
 -\delta_{K,k}\delta_{L,l}\delta_{M,m+1}\delta_{N,n}
+2\delta_{K,k}\delta_{L,l}\delta_{M,m}\delta_{N,n}
\end{eqnarray}
\begin{eqnarray}
\nonumber
 -\delta_{K,k}\delta_{L,l}\delta_{M,m}\delta_{N,n-1}
 -\delta_{K,k}\delta_{L,l}\delta_{M,m}\delta_{N,n+1}
)\Delta^{2}].
\end{eqnarray}
\begin{eqnarray}
\end{eqnarray}
\par
A set of eigenvalues $\eta^{a}$, associated with eigenvectors $x^{a}_{({\rm Q})\mu q}$ as well as $A^{a}_{({\rm Q})\mu q}$, is derived from a secular equation
\begin{eqnarray}
\sum_{q}M^{{\rm Q},{\rm Q}}_{pq}x^{a}_{({\rm Q})\mu q}
=\eta^{a} x^{a}_{({\rm Q})\mu p}.
\end{eqnarray}
The $v$-th eigenvector $x^{av}_{({\rm Q})\mu q}$, which will be shown to satisfy the secular equation, is given by
\begin{eqnarray}
\nonumber
x^{av}_{({\rm Q})\mu q}=x^{av}_{({\rm Q})\mu(K,L,M,N)}
\end{eqnarray}
\begin{eqnarray}
=\frac{1}{C_{\rm N}}
\sin (\vartheta_{K}j_{0}) \sin (\vartheta_{L}j_{1})
\sin (\vartheta_{M}j_{2}) \sin (\vartheta_{N}j_{3}),
\end{eqnarray}
where $C_{\rm N}$ is a normalization constant, and the
index
$j_{\mu}$ runs from $1$ to $N_{(\mu)}$ (the number of hypercube central points). The quantities $\vartheta_{K}$ (for $\mu=0$, for instance) are determined from a boundary condition. When the condition $\sin(\vartheta_{K}j_{0})=0$ for $K=N_{(0)}+1$ is imposed, we get
\begin{eqnarray}
\vartheta_{K}=\frac{K\pi}{N_{(0)}+1}.
\end{eqnarray}
\par
Setting $p=(k,l,m,n)$ and $q=(K,L,M,N)$, along with the use of the components for $A^{a}_{({\rm Q})\mu p}$ in $S^{(2)a}_{\rm Q,Q}$ expressed in terms of $\delta_{K,k^{\prime}}\delta_{L,l^{\prime}} \delta_{M,m^{\prime}}\delta_{N,n^{\prime}}$, the secular equation becomes (for example, for $\mu=0$)
\begin{eqnarray}
\nonumber
-\eta^{av} x^{av}_{({\rm Q})0(k,l,m,n)}+
\end{eqnarray}
\begin{eqnarray}
\nonumber
\frac{1}{2}[(2x^{av}_{({\rm Q})0(k,l,m,n)}-x^{av}_{({\rm Q})0(k-1,l,m,n)}
-x^{av}_{({\rm Q})0(k+1,l,m,n)})
\end{eqnarray}
\begin{eqnarray}
\nonumber
+(2x^{av}_{({\rm Q})0(k,l,m,n)}-x^{av}_{({\rm Q})0(k,l-1,m,n)}
-x^{av}_{({\rm Q})0(k,l+1,m,n)})
\end{eqnarray}
\begin{eqnarray}
\nonumber
+(2x^{av}_{({\rm Q})0(k,l,m,n)}-x^{av}_{({\rm Q})0(k,l,m-1,n)}
-x^{av}_{({\rm Q})0(k,l,m+1,n)})
\end{eqnarray}
\begin{eqnarray}
\nonumber
(2x^{av}_{({\rm Q})0(k,l,m,n)}-x^{av}_{({\rm Q})0(k,l,m,n-1)}
-x^{av}_{({\rm Q})0(k,l,m,n+1)})]
\end{eqnarray}
\begin{eqnarray}
\times\Delta^{2}=0.
\end{eqnarray}
The equation above is reduced to
\begin{eqnarray}
\nonumber
-\eta^{av}x^{av}_{({\rm Q})0 p}+[(1-c_{0})x^{av}_{({\rm Q})0 p}
+(1-c_{1})x^{av}_{({\rm Q})0p}
\end{eqnarray}
\begin{eqnarray}
+(1-c_{2})x^{av}_{({\rm Q})0 p}+(1-c_{3})x^{av}_{({\rm Q})0 p}]\Delta^{2}=0,
\end{eqnarray}
with 
\begin{eqnarray}
c_{\mu}=\cos (z_{\mu}),
z_{\mu}=\frac{j_{\mu}\pi}{N_{(\mu)}+1}.
\end{eqnarray}
(The summation convention is not applied to $(\mu)$ in parenthesis.) From the secular equation, we derive the eigenvalues
\begin{eqnarray}
\nonumber
\eta^{av}_{({\rm Q})(0)}=[(1-c_{0})+(1-c_{1})+(1-c_{2})+(1-c_{3})]
\Delta^{2},
\end{eqnarray}
\begin{eqnarray}
\end{eqnarray}
where the eigenvalues are positive. When the periodic boundary condition is imposed on the matrix elements, $c_{\mu}$ are replaced by
\begin{eqnarray}
c_{\mu}=\cos (z^{\prime}_{\mu}), z^{\prime}_{\mu}
=\frac{j_{\mu}(2\pi)}{N_{(\mu)}},
\end{eqnarray}
and zero eigenvalue appears. Thus, by imposing a non-periodic boundary condition, a zero-eigenvalue is avoided.
\par

\subsection{Wilson loop for quantum fluctuations}

Finally, we examine the contribution arising from quantum fluctuations to the
Wilson loop. Although $A^{a}_{\rm (C) \mu}=\lambda^{a}\tilde{A}_{\rm (C) \mu}$ for $1 \leq a \leq 3$, and $A^{a}_{\rm (C) \mu}=0$ for $a \geq 4$ in the classical case, this condition is not imposed on quantum fluctuations. From the aforementioned derivations, the action is given by
\begin{eqnarray}
S_{\rm Q}=\sum_{a}\sum_{p,q} M_{pq}
A^{a}_{{\rm (Q)}\mu p}A^{a}_{{\rm (Q)}\mu q}.
\end{eqnarray}
In the present scheme, the Wilson loop passes through the
points $x_{\mu p}$ (that is, $(t_{k},x_{l},y_{m},z_{n})$) and is expressed as
\begin{eqnarray}
\nonumber
W_{\rm Q}=
\end{eqnarray}
\begin{eqnarray}
\nonumber
{\rm Tr}\biggl \{
(I^{\prime (3)} W_{\rm C})
\frac{1}{Z_{\rm N}} \int D[A^{a}_{{\rm (Q)}\mu p}]
\exp(-S_{\rm Q}) \exp(C)
\biggr \},
\end{eqnarray}
\begin{eqnarray}
\end{eqnarray}
where
\begin{eqnarray}
Z_{\rm N}= \int D[A^{a}_{{\rm (Q)}\mu p}] \exp(-S_{\rm Q}),
\end{eqnarray}
\begin{eqnarray}
\nonumber
C=-\sum_{a}\sum_{p}
ig\oint dx_{\mu} A^{a}_{{\rm (Q)} \mu}T^{a}
\end{eqnarray}
\begin{eqnarray}
=-ig \sum_{a}\sum_{p} \beta_{\mu p}
A^{a}_{{\rm (Q)}\mu p}T^{a},
\end{eqnarray}
with
\begin{eqnarray}
\beta_{\mu p} = \oint dx_{\mu} \Omega^{4}_{p}.
\end{eqnarray}
In the classical part, 
\begin{eqnarray}
I^{\prime(3)}=
\left[\begin{array}{cc}
I^{(3)} &{\bf 0} \\
{\bf 0} &{\bf 0} \\
\end{array}\right],
\end{eqnarray}
where $I^{(3)}$ is the $SU(2)$ unit matrix and $I^{\prime(3)}$ is a matrix of ${\rm SU} (N)$ such as ${\rm SU}(3)$.
\par
As was described in the previous section, the matrix $M_{pq}$ is diagonalized by the transformation $A^{\prime a}_{{\rm (Q)}\mu p}=\sum_{q}R_{pq}A^{a}_{{\rm (Q)}\mu q}$ and the action has the form
\begin{eqnarray}
S_{\rm Q}=\sum_{a}\sum_{p} \eta^{a}_{(\mu)p}
(A^{\prime a}_{{\rm (Q)}\mu p})^{2},
\end{eqnarray}
where $\eta^{a}_{(\mu)p}$ is the eigenvalue associated with $A^{\prime a}_{{\rm (Q)}\mu p}$. (The summation convention is not applied to $(\mu)$ in parenthesis.) Then, the Wilson loop is written as
\begin{eqnarray}
\nonumber
W_{\rm Q}={\rm Tr}
\biggl \{
(I^{\prime (3)} W_{\rm C}) \frac{1}{Z_{\rm N}} \int D[A^{\prime a}_{{\rm (Q)}\mu p}]
\end{eqnarray}
\begin{eqnarray}
\nonumber
\times
\exp(-i \sum_{a} \sum_{p} B_{\mu p}^{\prime a}
A_{{\rm (Q)}\mu p}^{\prime a})
\end{eqnarray}
\begin{eqnarray}
\times
\exp(- \sum_{a} \sum_{p}
\eta^{a}_{(\mu)p}(A^{\prime a}_{{\rm (Q)}\mu p})^{2})
\biggr \},
\end{eqnarray}
where
\begin{eqnarray}
B_{\mu p}^{\prime a}= g \sum_{q}
\beta_{\mu q}R^{-1}_{qp} T^{a}.
\end{eqnarray}
\par\
Because odd terms with respect to $A^{\prime a}_{{\rm (Q)}\mu p}$ vanish in the Gaussian integral independent of the matrices $T^{a}$, the Wilson loop is given by
\begin{eqnarray}
\nonumber
W_{\rm Q}={\rm Tr}
\biggl \{
(I^{\prime (3)} W_{\rm C} ) \frac{1}{Z_{\rm N}}\Pi_{a}\Pi_{p}\int
dA^{\prime a}_{{\rm (Q)}\mu p}
\end{eqnarray}
\begin{eqnarray}
\nonumber
\times
\cos(B^{\prime a}_{\mu p} A^{\prime a}_{{\rm (Q)}\mu p})
\exp[-\eta^{a}_{(\mu)p}(A^{\prime a}_{{\rm (Q)}\mu p})^{2}]
\biggr \}
\end{eqnarray}
\begin{eqnarray}
\nonumber
={\rm Tr}
\biggl \{
(I^{\prime (3)} W_{\rm C})\frac{1}{Z_{\rm N}} \Pi_{a}\Pi_{p}
( \frac{\pi}{ \eta^{a}_{(\mu)p} } )^{1/2}
\exp[-\frac{(B^{\prime a}_{\mu p})^{2}}
{4\eta^{a}_{(\mu)p}}]
\biggr \}.
\end{eqnarray}
\begin{eqnarray}
\end{eqnarray}
The normalization constant $Z_{\rm N}$ is denoted as 
\begin{eqnarray}
Z_{\rm N}=\Pi_{a}\Pi_{p} ( \frac{\pi}{ \eta^{a}_{(\mu)p} })^{1/2}.
\end{eqnarray}
Therefore, by writing as ${B}^{\prime a}_{\mu p}= \tilde{B}^{\prime}_{\mu p}T^{a}$, we get the following Wilson loop
\begin{eqnarray}
\nonumber
W_{\rm Q}={\rm Tr}
\biggl \{
(I^{\prime (3)} W_{\rm C} ) \exp[-\sum_{a} \sum_{p}
\frac{(\tilde{B}^{\prime}_{\mu p})^{2}}
{4\eta^{a}_{(\mu)p}}T^{a}T^{a}]
\biggr \}.
\end{eqnarray}
\begin{eqnarray}
\end{eqnarray}
In the relation above, $\sum_{a} T^{a}T^{a}$ is proportional to the unit matrix, and the quantum contribution to the Wilson loop from the non-Abelian gauge field mentioned above has the same form as that from the Abelian gauge field.
\par
The Wilson loop is then given by
\begin{eqnarray}
W
_{\rm Q}
=W_{\rm C}\exp[-(t_{2}-t_{1})V_{\rm C}(x_{2}-x_{1})],
\end{eqnarray}
where $V_{\rm C}$ denotes a Coulomb potential. Hence, (by dropping the constant) we have the quantum potential in addition to the classical potential
\begin{eqnarray}
\nonumber
V(x_{2}-x_{1})=-\frac{1}{t_{2}-t_{1}}\ln W
_{\rm Q}
\end{eqnarray}
\begin{eqnarray}
=\frac{a_{\rm c}}{2}(x_{2}-x_{1})+V_{\rm C}(x_{2}-x_{1}).
\end{eqnarray}
The analytical confining potential derived above is composed of a linear term + Coulomb term. Previously, a confining potential composed of linear and Coulomb terms was obtained numerically using the Wilson's lattice gauge theory. In addition to fermion confinement, this mechanism
also works 
for pure non-Abelian fields. When two field sources with the opposite signs are created due to self-interaction, classical fields
are
shifted from zero field, and
generate
local
mass
(energy gap) for quantum fluctuations.
\par
The contribution of quantum fluctuations to the Polyakov line is also independent of the gauge group, and the new phenomenon beyond the Abelian case does not occur in this case.
\par

\section{Conclusions}

We have presented an example of an analytic function for classical vector potentials of Yang-Mills fields for a heavy fermion-antifermion pair, which leads to fermion confinement. The string tension of the linear potential between the particles was derived and its physical implications were provided. Quantum fluctuations around the classical field were described using real spacetime basis functions in the path integral scheme. The quantum fluctuations dealt with a Coulomb term in addition to the linear potential caused by the classical fields between particles.
\par

\end{multicols}

\end{document}